%% file: main.tex
\documentclass[acmsmall,screen,nonacm]{acmart}

\input{setup/00-preamble}
\input{setup/03-macros}

\begin{document}

\title{Structural Code Search using Natural Language Queries}
%\miryung{consider retitle to fit PLDI: how about this? Automated Translation to Structural Code Queries from Natural Language Description? }
%\pranav{How about: Structural Code Search using Natural Language Queries?} 
%\miryung{I like the new title}
\input{setup/01-authors}

\input{sections/00-abstract}

\input{setup/02-metadata}

\maketitle

\input{sections/10-introduction}

\input{sections/11-background}

\input{sections/12-approach}

\input{sections/12-implementation}
\input{sections/13-evaluation}
\input{sections/15-threats-to-validity}

\input{sections/16-related-work}
\input{sections/17-conclusion}

% \begin{acks}
% \end{acks}

\bibliographystyle{ACM-Reference-Format}
\bibliography{main}

\end{document}

%% file: setup/00-preamble.tex
\usepackage{enumitem}
\usepackage{booktabs}
\usepackage{listings}
\usepackage{xcolor}
\usepackage{hyperref}
\usepackage{algpseudocode}
\usepackage{algorithm}
\usepackage{subcaption}
\usepackage{multirow}
\usepackage{tabularx}
\usepackage{tcolorbox}

\lstnewenvironment{gql}
  {\lstset{language=Java,
    basicstyle=\footnotesize,% basic font setting
    }%
  }
  {}

\setcopyright{acmlicensed}
\copyrightyear{2018}
\acmYear{2018}
\acmDOI{XXXXXXX.XXXXXXX}

\acmConference[PLDI '25]{Programming Language Design and Implementation}{June 16--20,
  2025}{Seoul, South Korea}
\acmISBN{978-1-4503-XXXX-X/18/06}

%% file: setup/03-macros.tex
\newcommand{\code}[1]{\texttt{\small #1}}

%% file: setup/01-authors.tex
% \author{Ben Limpanukorn}
% \authornote{Work done during internship at AWS.}
% \email{blimpan@cs.ucla.edu}
% % \orcid{1234-5678-9012}
% \affiliation{%

\author{Ben Limpanukorn}
\authornote{Work done during internship at Amazon Web Services.}
\affiliation{%
  \institution{University of California, Los Angeles}
  % \city{Hekla}
  \country{USA}}
\email{blimpan@cs.ucla.edu}

\author{Yanjun Wang}
\affiliation{%
  \institution{Amazon Web Services}
  % \city{Hekla}
  \country{USA}}
\email{yanjunw@amazon.com}

\author{Zach Patterson}
\affiliation{%
  \institution{Amazon Web Services}
  % \city{Hekla}
  \country{USA}}
\email{pattzac@amazon.com}

\author{Pranav Garg}
\affiliation{%
  \institution{Amazon Web Services}
  % \city{Hekla}
  \country{USA}}
\email{prangarg@amazon.com}

\author{Murali Krishna Ramanathan}
\affiliation{%
  \institution{Amazon Web Services}
  % \city{Hekla}
  \country{USA}}
\email{mkraman@amazon.com}

\author{Xiaofei Ma}
\affiliation{%
  \institution{Amazon Web Services}
  % \city{Hekla}
  \country{USA}}
\email{xiaofeim@amazon.com}

\author{Anoop Deoras}
\affiliation{%
  \institution{Amazon Web Services}
  % \city{Hekla}
  \country{USA}}
\email{adeoras@amazon.com}

\author{Miryung Kim}
\authornote{Miryung Kim holds concurrent appointments as an Amazon Scholar and as a Professor of Computer Science at the University of California, Los Angeles. This paper describes work performed at Amazon.}
\affiliation{%
  \institution{Amazon Web Services}
  % \city{Hekla}
  \country{USA}}
\email{miryung@amazon.com}

% \affiliation{%

% \affiliation{%

% \affiliation{%

% \affiliation{%

% \affiliation{%

% \affiliation{%

\renewcommand{\shortauthors}{Limpanukorn et al.}

%% file: sections/00-abstract.tex
\begin{abstract}
Searching code is a common task that developers perform to understand APIs, learn common code patterns, and navigate code.
Currently, developers most commonly search using keywords and regular expressions that are easy to use and widely available.
Beyond keywords and regular expressions, structural code search tools allow developers to search for code based on its syntactic structure. This has numerous applications ranging from bug finding to systematically refactoring code~\cite{coccinelle-origins}. 
However, these structural code search tools operate on queries  expressed in domain-specific languages (DSL) that can be difficult to learn and write.
We propose to allow developers to use natural language to search for code structurally.
Expressing queries in natural language provides an intuitive way to search for code and lowers the barrier to entry.

In this work, we develop a novel general approach that combines the reasoning capabilities of an LLM to interpret natural language search queries with the power of structural search tools to efficiently and accurately retrieve relevant code.
We then instantiate this approach for two structural code search DSLs: Semgrep and GQL.
In our evaluation, we construct a new benchmark for structural code search consisting of 400 queries over 10 Java projects.
We show that our approach for structural code search based on translating NL queries to DSL queries using an LLM is effective and robust, achieving a high precision and
recall ranging from 55\% - 70\%. Further, our approach significantly outperforms baselines based
on semantic code search and LLM retrievals by up to 57\% and 14\% on F1 scores.

% Our approach achieves F1 score of 58\% using GQL and 70\% using Semgrep, outperforming LLM and similarity search baselines by up to 14\% and 57\% respectively.% and outperforms other baselines by $Y\times$.
%\pranav{1/ emphasize using NL in intuitive and lowers the barrier of entry for developers 2/ We develop a general approach and instantiate it with  two DSLs for structural code search: Semgrep and GQL. }

\end{abstract}

%% file: setup/02-metadata.tex


\ccsdesc[500]{Information systems~Search interfaces}
\ccsdesc[500]{Information systems~Query representation}
\ccsdesc[500]{Information systems~Question answering}
\ccsdesc[500]{Information systems~Structured text search}

\keywords{Structural Code Search, Code Search, LLM, RAG, GQL, Semgrep}


%% file: sections/10-introduction.tex
\section{Introduction}

Searching code is one of the most common capabilities developers use in their day-to-day work. Developers use code search to understand APIs, learn common code patterns, and navigate code~\cite{sadowski-code-search}.
Search using keywords and regular expressions are language-agnostic, easy to use and are included in all major code platforms.
Developers, though, have use for code search that goes beyond keywords and regular expression search.
Structural code search allows developers to search for code based on its syntactic structures with applications ranging from bug finding to systematic code refactoring~\cite{coccinelle-origins}. Structural code search is present in all major IDEs~\cite{intellij, vscode} and also in standalone search tools such as SourceGraph~\cite{sourcegraph}.
However, these structural search tools require the users to search using queries expressed in domain-specific languages (DSL) that can be difficult to learn and write~\cite{pradel-code-search-survey}.
This limits developers from using these tools in their daily workflows.
In this work, we propose that natural language be used to search for code structurally. 
This provides an intuitive way for developers to search for code without learning a new DSL.

Existing approaches for code search using natural language includes semantic code search where search is performed by computing a semantic embedding of the natural language query and finding code chunks whose embeddings have the highest similarity~\cite{codesearchnet, neuralcodesearch, cambronero-code-search}.
However, similarity search methods are imprecise when answering \emph{structural} code search queries which contain complex constraints that are not precisely captured by the embeddings alone.
For example, a structural query to \textit{"find all calls to foo() that take a String as an argument"} would require the search engine that can reason over type information and scope in the program.
An alternative approach is presented by large language models (LLMs) that can be prompted to directly answer search queries 
by retrieving relevant code from the input code context~\cite{repoqa}. However, LLMs are highly inefficient in terms of the number of input and output tokens, latency and cost. This makes pure LLM based structural code search systems nonviable for search applications involving large code bases. %over larger code bases and suffers from poor recall with larger context sizes.

In this work, we describe a novel approach that enables developers to search code structurally from natural language queries and present a benchmark for evaluating such code search tools.
Our approach leverages the reasoning capabilities of an LLM with retrieval augmented generation (RAG) to translate structural code search queries from natural language (NL) to the DSL of a structural code search tool. 
%Currently, no dataset or benchmark of structural code search queries exists.
We achieve this by proposing an algorithm that synthetically generates examples of paired (NL, DSL) structural code search queries. 
%\pranav{say we are working to release this dataset as open source}
We use this algorithm to generate a training set for RAG and to construct a benchmark to evaluate structural code search tools. This is the first natural language based structural code search dataset and we are currently working to open source this dataset for public use. 

Our evaluations show that our approach for structural code search based on translating NL queries to DSL queries using an LLM is effective and robust, achieving a high precision and recall ranging from 55\% - 70\%.  
Further, our approach significantly outperforms baselines based on semantic code search and LLM retrievals by up to 57\% and 14\% on F1 score. 
Finally, we report ablation studies that motivate different components in our proposed approach.

%We are currently working to open source both the training and benchmark datasets.
%We are working to 

%\pranav{do we need to address the question that though NL can be vague is the most intuitive way to express structural code search queries?}

In this work, we present the following key contributions:
\begin{enumerate}
    \item A generic algorithm for automatically generating structural code search queries in natural language and domain-specific languages. We instantiate this algorithm with two structural code search DSLs: Guru Query Language (GQL)~\cite{mukherjee-gql} and Semgrep~\cite{semgrep}.
    \item A set of benchmarks to evaluate structural code search tools consisting of 400 queries over 10 Java projects.
    \item A novel approach to enable developers to search for code structurally using queries expressed in natural language that achieves a high precision and recall ranging from 55\% - 70\% on our GQL-derived and Semgrep-derived benchmark datasets.
\end{enumerate}

This paper is structured as follows: We first provide background and a motivating example in Section~\ref{sec:background}. We then describe our approach in general terms in Section~\ref{sec:approach}, and instantiate our approach for Semgrep and GQL in Section~\ref{sec:implementation}. We discuss our evaluation setup in Section~\ref{sec:eval-setup} and present our results in Section~\ref{sec:evaluation}. We discuss related work in Section~\ref{sec:related-work} and conclude in Section~\ref{sec:conc}.

%% file: sections/11-background.tex
\section{Background and Motivating Example}
\label{sec:background}

% \begin{table}[h]
%     \centering
%     \begin{tabular}{c|cccc}
%     \toprule
%          DSL&   Style&Data-flow &Control-flow&Taint Tracking\\
%          \midrule
%          GQL & 
%      Imperative&Yes  &Yes&Yes\\
%  Semgrep & Declarative, Code-like&No  &Yes&Yes\\
%  CodeQL & Declarative, SQL-like&Yes  &Yes&Yes\\
%  SourceGraph/Comby & Declarative, Code-like& No &Yes&No\\
%  \bottomrule
%  \end{tabular}
%     \caption{Comparison of Structural Code Search DSLs}
%     \label{tab:my_label}
% \end{table}

There are a number of DSLs one may use for structural code search that vary in their expressivity. In Table~\ref{table:background}, we list a few popular DSLs-- Comby\cite{comby}, Semgrep~\cite{semgrep}, CodeQL~\cite{codeql} and GQL~\cite{mukherjee-gql}. To give a flavor of these DSLs, we include in the table a partial listing of the query predicates, language constructs, expressivity characteristics and an example query. 
As a motivating example, consider a scenario where a developer wants to identify  \code{String} concatenations in \code{for} loops in Java code. Such concatenation operations are inefficient and the best practice is to use a \code{StringBuilder} object to construct such \code{String}s. Presently, developers need to first learn the DSL syntax to precisely express this pattern and then they may search for such concatenation operations in their code (Table~\ref{table:background} lists queries that check a variant of this property in different DSLs). However, with natural language based structural code search, developers can intuitively search for such concatenation operations by issuing a simple NL query: \textit{"Find all cases where a String object is used by an add operator within a for loop"}. This alleviates the burden of learning the DSL completely and broadens the appeal and applications of structural code search amongst developers. 
% By eliminating the need to learn such DSLs, and instead enabling developers to express their intent in natural language, we lower the entry barrier for wider adoption of these tools in day-to-day software development beyond a fixed set of pre-written rules.

\begin{table}[ht!]
\caption{Popular DSLs for structural code search}
\label{table:background}
\begin{tabularx}{\textwidth}{llll}
\toprule
Query Language &
  Searchable predicates &
  Language constructs &
  Expressivity \\ \midrule

\multirow{2}{*}{Comby} &
  \begin{tabular}[c]{l}
        \code{for(...) \{ ... \} }, \\ 
        \code{:[hole] = ... }, \\
        \code{:[hole\textasciitilde[a-z]](...)}, \dots
    \end{tabular} &
  \begin{tabular}[c]{l} 
    \code{where [:a] == [:b]}, \\ 
    \code{where match [:a] \{ \}}, \\
    \dots
  \end{tabular} &  
  \begin{tabular}[c]{l}
        Data-flow: approx.\\ 
        Control-flow: Yes \\ 
        Inter-proc: No
  \end{tabular} \\
  \cmidrule{2-4}
 &
\multicolumn{3}{l}{
    \begin{minipage}{0.77\linewidth}  
    NL Query:
        \textit{"Find all cases where value returned by \texttt{\small Integer.toString} is used by an add operation inside a for loop"} \\
        DSL Query: \\
        \code{for( ... ) \{:[X] += Integer.toString(...);\}}
    \end{minipage}    
    } \\
\midrule
\multirow{2}{*}{Semgrep} &  
  \begin{tabular}[c]{l}
        \code{for(...) \{ ... \} }, \\ 
        \code{\$X = ... }, \\
        \code{\$Call(...)}, \\
        \code{...}
    \end{tabular} &
  \begin{tabular}[c]{l} 
    \code{pattern}, \\ 
    \code{pattern-either}, \\ 
    \code{pattern-not},  \\ 
    \code{pattern-inside}, \dots
  \end{tabular} &  
  \begin{tabular}[c]{l}
        Data-flow: approx.\\ 
        Control-flow: Yes \\ 
        Inter-proc: No
  \end{tabular} \\
  \cmidrule{2-4}
 &
\multicolumn{3}{l}{
    \begin{minipage}{0.77\linewidth}  
    NL Query:
        \textit{"Find all cases where value returned by \texttt{\small Integer.toString} is used by an add operation inside a for loop"} \\
        DSL Query: \\
        \code{pattern: \$X +=   Integer.toString(...);} \\        
        \code{pattern-inside: for( ... ) \{ ... \}}     
    \end{minipage}    
    } \\
\midrule
\multirow{2}{*}{CodeQL} &  
  \begin{tabular}[c]{l}
        \code{hasName(...)}, \\ 
        \code{hasQualifiedName(...)} \\
        \code{getIntValue(...) = ...} \\
        \dots
    \end{tabular} &
  \begin{tabular}[c]{l} 
    \code{or}, \code{and}, \code{not}, \\ 
    \code{forall}, \code{exists}, \\
    \code{if then else} \\
    \dots
  \end{tabular} &  
  \begin{tabular}[c]{l}
        Data-flow: Yes \\ 
        Control-flow: Yes \\ 
        Inter-proc: No
  \end{tabular} \\
  \cmidrule{2-4}
 &
\multicolumn{3}{l}{
    \begin{minipage}{0.77\linewidth}  
    NL Query:
        \textit{"Find all cases where \texttt{equals} is called on an empty string."} \\
        DSL Query: \\
\code{from MethodAccess ma where} \\ 
\code{    ma.getMethod().hasName("equals") and} \\
\code{    ma.getArgument(0).(StringLiteral).getValue() = ""} \\
\code{select ma, "Matched"}
    \end{minipage}    
    } \\
\midrule
\multirow{2}{*}{GQL} &
  \begin{tabular}[c]{l} 
        \code{withMethodCallFilter}, \\  
        \code{withDataByTypeFilter},\\ 
        %\code{withActionFilter}, \\ 
        \code{withDataDependencies-} \\ 
        \code{  Transform}, \dots 
    \end{tabular} &    
  \begin{tabular}[c]{l} 
        \code{withAnyOf}, \\ 
        \code{withAllOf},\\  
        \code{withNegationOf}, ... 
    \end{tabular} &
    \begin{tabular}[c]{l} 
        Dataflow: Yes\\ 
        Control-flow: Yes\\  
        Inter-proc: Yes
    \end{tabular} \\
    \cmidrule{2-4}
 &
 \multicolumn{3}{l}{
    \begin{minipage}{0.77\linewidth}  
    NL Query:
        \textit{"Find all cases where a String object is used by an add operator within a for loop"} \\
    DSL Query: \\
        \code{new CustomRule.Builder()} \\        
        \code{.withControlFilter("FOR\_STATEMENT") // Match for loops} \\
        \code{.withContextNodesTransform(ContextKind.LOOP) // Transform to body} \\
        \code{.withActionFilter("$\backslash \backslash$+") // Filter for the add operator} \\
        \code{.withArgumentTransform(ArgumentPredicate.ANY\_ARGUMENT)} \\
        \code{.withDataByTypeFilter(true, "String") // Match String objects} \\
        \code{.check().build()}        
    \end{minipage}    
    } \\
\bottomrule    
\end{tabularx}
\end{table}

%\pranav{Brief description of why the target DSL may influence NL queries.}
%\ben{adding to next paragraph:}

While our approach for structural code search is DSL-agnostic, in this paper, we instantiate the approach with two DSLs: Semgrep and GQL. 
The choice of DSL influences the space of expressible natural language queries as each language supports slightly different predicates and features.
For instance, GQL provides predicates for full data-flow analysis which enables queries such as \textit{"Find all if statements that depend on a variable that is also used by foo()"} which cannot be precisely expressed in Semgrep. On the other hand, Semgrep allows patterns over class and method declarations that is not supported in GQL.

With this overview, let us now briefly describe GQL and Semgrep in more detail in the following subsections. 

\subsection{Semgrep}

Semgrep is a declarative language that allows developers to match generalized patterns over abstract syntax trees (ASTs). An appealing feature of Semgrep is that AST patterns are expressed as code snippets in the target language (e.g., Java), which makes it quite developer friendly. These AST patterns can be generalized by augmenting the target language syntax with meta-variables (e.g., \$X in the motivating example in Table~\ref{table:background}) and ellipses (\code{...}). Further, Semgrep augments structural matching over ASTs with semantic analyses such as type inference and constant propagation. As listed in Table~\ref{table:background}, Semgrep syntax allows composing multiple patterns using conjunctions (\code{patterns}), disjunctions (\code{pattern-either}) and negations (\code{pattern-not}). For the motivating example, the \code{pattern} construct in Semgrep query searches for a \code{+=} operation with \code{Integer.toString} method call as the right operand. Further, the \code{pattern-inside} construct in the query checks that the matching \code{+=} operation must reside inside a \code{for} loop.

\subsection{GQL}

Guru Query Language (GQL)~\cite{mukherjee-gql} is a proprietary Java-based DSL from Amazon that developers can use to express code patterns over a program dependence representation called MuGraph~\cite{mukherjee-gql}. As the name suggests, MuGraph is a graph representation where nodes represent actions (e.g., function calls, operations, control statements) or data (e.g., objects and variables), and edges represent relationships, such as control dependence or data flow, between the nodes. GQL exposes a Java Builder pattern that developers can use to chain together different filter operations (project to a subset of MuGraph nodes that satisfy a property) or transforms (select nodes related to a given node via a MuGraph edge) to express complex code patterns such as \code{String} concatenations in the motivating example. For this example, the GQL query in Table~\ref{table:background} first matches all \code{for} statements with a control filter. It then transforms to all nodes \textit{inside} the body of these \code{for} statements, followed by a filter operation that selects only the subset of these nodes that correspond to the \code{+} operation. Subsequently, the query transforms the \code{+} node to its arguments and checks using \code{withDataByTypeFilter} if any of the argument is of type \code{String}.

% Comby
% \ben{\url{https://sourcegraph.com/blog/how-to-search-with-sourcegraph-using-structural-patterns}
% \begin{itemize}
%     \item very similar to Semgrep (minus metavariables)
%     \item no taint-analysis mode
%     \item based on Comby syntax
% \end{itemize}
% }

% \begin{figure}[h]
%     \centering
%     \begin{subfigure}{0.45\textwidth}
%     \includegraphics[width=\textwidth]{figures/MotivatingMUGraph.png}
%     \caption{A partial MU graph that corresponds to the code snippet in Listing \ref{lst:code-to-match}.}
%     \end{subfigure}
%     \hfill
%     \begin{subfigure}{0.49\textwidth}
%     \begin{lstlisting}[language=Java, captionpos=b, label={lst:code-to-match}, caption={An example of an inefficient String concatenation within a for loop.}]
% String a;
% for (int i = 0; i < limit; i++) {
%     a += Integer.toString(number);
% }
%     \end{lstlisting}
%     \end{subfigure}
%     \caption{An example code snippet to be matched by a structural code query and its corresponding MU graph. Solid lines in the graph indicate control-flow relations and dotted lines indicate data-flow relations. The "dummy\_" node represents the value returned by "Integer.toString()" and the "\_phi\_" node represents a converging data-flow (i.e. String \texttt{a} is concatenated with itself).}
%     \label{fig:mu-graph}
% \end{figure}

%% file: sections/12-approach.tex
\section{Approach for Structural Code Search}
\label{sec:approach}

\begin{figure}[h]
    \centering
    \includegraphics[width=0.8\textwidth]{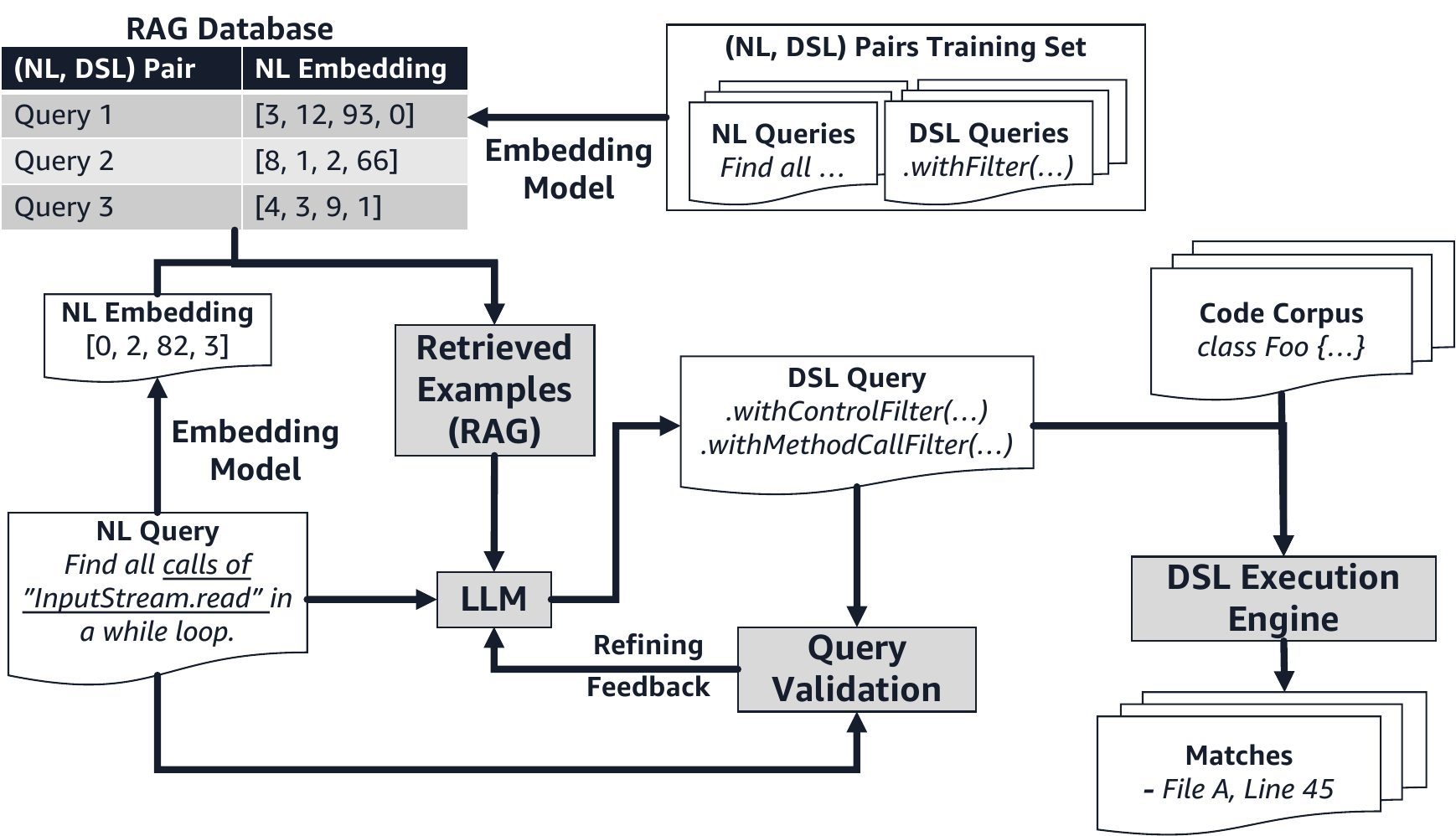}
    \caption{Proposed approach for structural code search for queries posed in natural language. %\pranav{We should make the figure generic and not tied to GQL.}
    }    
    \label{fig:nl-to-dsl-pipeline}
\end{figure}

In this section we describe a generic approach for structural code search using natural language queries. We then instantiate this approach for two different structural code search DSLs -- GQL and Semgrep, in Section~\ref{sec:implementation}. 

We describe an overview of our approach in Figure~\ref{fig:nl-to-dsl-pipeline}. Given a structural code search query in natural language, our approach uses a large language model (LLM) to translate it into a search query expressed in a DSL of choice. This search query is then executed on code corpus using the execution engine of the DSL to retrieve the matching code. In general, translating the natural language query into a DSL accurately is hard for lack of a large training corpus and we use a retrieval-augmented generation (RAG)~\cite{lewis-rag} setup for this translation. We instantiate RAG with few-shot pairs of (NL, DSL) queries that we generate 
using a systematic generation of DSL queries and an LLM to pair them with corresponding NL queries (described in Section~\ref{sec:query-enumeration} and~\ref{sec:deriving-nl-dsl-pairs}). 
If the DSL query generated by the LLM is malformed or incorrect, we use an LLM based query refinement to fix the query (described in Section~\ref{sec:error-feedback}). 

With this overview, we describe each component of our approach in  the subsections below.

% To provide these examples, we enumerate a training dataset of (NL, DSL) structural code query pairs.
% When prompted to translate an NL query, the (NL, DSL) training pairs with the highest embedding similarity are retrieved and provided to the LLM.

%\ben{TODO: insert prompt figure here}

\input{sections/approach/1-query-enumeration}

\input{sections/approach/2-deriving-nl-dsl-pairs}
\input{sections/approach/3-rag}
\input{sections/approach/4-error-feedback}

%% file: sections/approach/1-query-enumeration.tex
\subsection{DSL Query Generation}
\label{sec:query-enumeration}

%\pranav{Overall algorithm to search using the generated NL-DSL pairs in a RAG setting should also come as a subsection in this section.}

To build a dataset of structural code search queries, we systematically generate search queries that match instances of code constructs from real-world software projects. We describe the algorithm in Algorithm~\ref{alg:dsl-generate}. 
Given a corpus of source code $T$, the algorithm returns a set of structural code search queries $Q$ in the chosen DSL.
Each search query targets a particular instance of a code construct $t \in T$,  
%$T \in \bb{T}$ represents a source file and is a set of code constructs in said source file.
where each code construct $t$ consists of a construct type (e.g. Literal, If Statement, Method Call, etc.) and a code span denoting its start and end location. Note, in this section, we describe the generation algorithm in general terms and provide instantiations for various sub-routines called in the algorithm for the GQL and Semgrep languages in Section \ref{sec:implementation}.

%Each query is then translated to a natural language form by prompting an LLM with an automatically derived description of the query.
During generation, we bias the selection of predicates to achieve a near-uniform distribution over a pre-defined set of code construct types.
This ensures the dataset contains a high diversity of queries and adequate coverage over the chosen constructs. 

%\pranav{mention we want to generate set of queries that have a high diversity and in the ideal case a near-uniform distribution over different code constructs. Also mention that we describe the algorithm in general terms and then provide two instantiations of this algorithm for GQL and Semgrep in Section 4}

%Let $\bb{E} = \{\text{Literal}, \text{If Statement}, \text{Method Call}, \dots\}$ be the set of code constructs that can be represented in the DSL.
%Let $\bb{P}$ be the set of available predicates in the DSL.

To generate a query, the algorithm first selects a code construct $t$ from the corpus $T$ (line 3).
Then, the query is initialized (line 4) and verified to ensure that it  matches the target code construct (lines 5-7).
Depending on the implementation of the \textsc{Init} method, the query at this point may have  a complexity higher or lower than the desired complexity. To meet the desired complexity goal, the algorithm  iteratively generalizes or specializes this query until the complexity of the updated query falls within the complexity targets ($c_{min}, c_{max}$) (lines 9-10). In each such iteration, the query is re-verified to ensure that it continues to match the-- possibly updated, code construct.
% \pranav{for semgrep, since we do not change target code construct across iterations how does sampling work?}
% \ben{for semgrep, there are only two opportunities to change the distribution: when the target code construct is first selected, and when a AST node is chosen to be "generalized"---replaced with a metavariable or an ellipsis.}

\subsubsection{Biasing generations to a more uniform distribution over code construct types}
During query generation, we track the frequency of each code construct type in the previously generated queries. A list of types of code constructs $E$ we have considered is shown in Figure~\ref{fig:benchmark-stats}. %(\pranav{we need to list all the code constructs someplace}).
This frequency information is used in the \textsc{WeightedSample}, \textsc{Specialize} and \textsc{Generalize} functions to up-sample code constructs that appear less frequently in previously generated queries. 
In particular, \textsc{WeightedSample} randomly selects a code construct instance $t$ of type $e$ with probability $1/(1+c_e)$ where $c_e$ is the number of instances of the construct type $e$ in $Q$. 
This biases the generations towards a more uniform distribution of code construct types in the resulting set of queries $Q$.

\begin{algorithm}[h]
\caption{Structural Code Search Query Enumeration Algorithm %\pranav{I think t will be determined with weighted sampling inside the while loop at line 9 in the algorithm again? Does t need to be passed in both Specialize and Generalize?}
}
\label{alg:dsl-generate}
\begin{algorithmic}[1]
\Require    
    \item[]
        \begin{itemize}
            \item $T \gets$ a code corpus.
            \item $n_Q \gets$ the number of queries to generate.
            \item $c_{min}, c_{max} \gets$ the minimum/maximum complexity of the query.
        \end{itemize}
\Ensure
    \item[] 
        \begin{itemize}
            \item $Q \gets$ a set of structural code search queries.
        \end{itemize}
\State $Q \gets \{\}$
\While{$|Q| < n_Q$}
    \State $t \gets \textsc{WeightedSample}(T, Q)$\Comment{Select a code construct.}
    \State $q \gets \textsc{Init}(t)$
    \If{$t \not\in \textsc{Execute}(q, T)$}
        \State \textbf{continue} $\hookleftarrow$ \Comment{Verify that the initialized query matches the target construct.}
    \EndIf
    \While{$\textsc{Complexity}(q) < c_{min} \lor \textsc{Complexity}(q) > c_{max}$}
        \State \textbf{if} {$\textsc{Complexity}(q) < c_{min}$} \textbf{then} $q', t' \gets \textsc{Specialize}(q, t, T, Q)$
        \State \textbf{if} {$\textsc{Complexity}(q) > c_{max}$} \textbf{then} $q' \gets \textsc{Generalize}(q, Q); t' \gets t$
        \If{$t' \in \textsc{Execute}(q', T)$}
            \State $q \gets q'$ \Comment{Verify that the modified query matches the updated target construct.}
            \State $t \gets t'$
        \Else
            \State \textbf{continue} $\hookleftarrow$
        \EndIf
    \EndWhile
    \State $Q \gets Q \cup \{q\}$
\EndWhile
\end{algorithmic}
\end{algorithm}

\subsubsection{Query Specialization and Generalization} 
The purpose of the \textsc{Specialize} and \textsc{Generalize} functions is to refine a query $q$ until the target complexity is reached. 
We define the $\textsc{Complexity}$ function to compute the complexity of $q$ as the number of distinct code constructs on which $q$ conditions.
As a simple example, complexity of a query $q$: "\textit{Find all calls to \texttt{foo()} controlled by an if statement}",  is two as it comprises two distinct code constructs--  the method call \code{foo()} and the \code{if} statement.

Given a query $q$, target code construct $t$ in the larger code context $T$, and set of previously generated queries $Q$, the \textsc{Specialize} function 
modifies the query by adding clauses or predicates to more precisely match an instance of a code construct $t' \in T$. 
Note that $t'$ may be a different code construct than $t$.
%(\pranav{what is t? is it the construct that needs to be used for specialization or generalization?}).
\textsc{Specialize} achieves the effect of increasing the complexity of the query as the 
returned query additionally checks for a match on $t'$. 
On the other hand, \textsc{Generalize} updates a query $q$ by removing clauses or predicates from $q$. This reduces the complexity of the query as it does not need to match on the removed clauses / predicates.

%\begin{algorithm}
%\caption{$\textsc{Verify}(q, t, T)$: A function that verifies query $q$ matches construct $t$ in source file $T$}
%\label{alg:gql-generate}
%\begin{algorithmic}[1]
%\Require    
%    \item[]
%        \begin{itemize}
%            \item $q \gets$ a structural code search query.
%            \item $t \gets$ an instance of a code construct in $T$.
%            \item $T \gets$ a source code file.
%        \end{itemize}
%\Ensure
%    \item[] 
%        \begin{itemize}
%            \item Return \textbf{true} if executing $q$ on $T$ returns $t$
%        \end{itemize}
%\State $R \gets \textsc{Execute}(q, T)$
%\If{$t \in R$}
%    \State \textbf{return true} $\hookleftarrow$
%\Else
%    \State \textbf{return false} $\hookleftarrow$
%\EndIf
%\end{algorithmic}
%\end{algorithm}

%% file: sections/approach/2-deriving-nl-dsl-pairs.tex
\subsection{Pairing the DSL query with NL query}
\label{sec:deriving-nl-dsl-pairs}

\begin{figure}[h]
    \centering
    \includegraphics[width=0.95\textwidth]{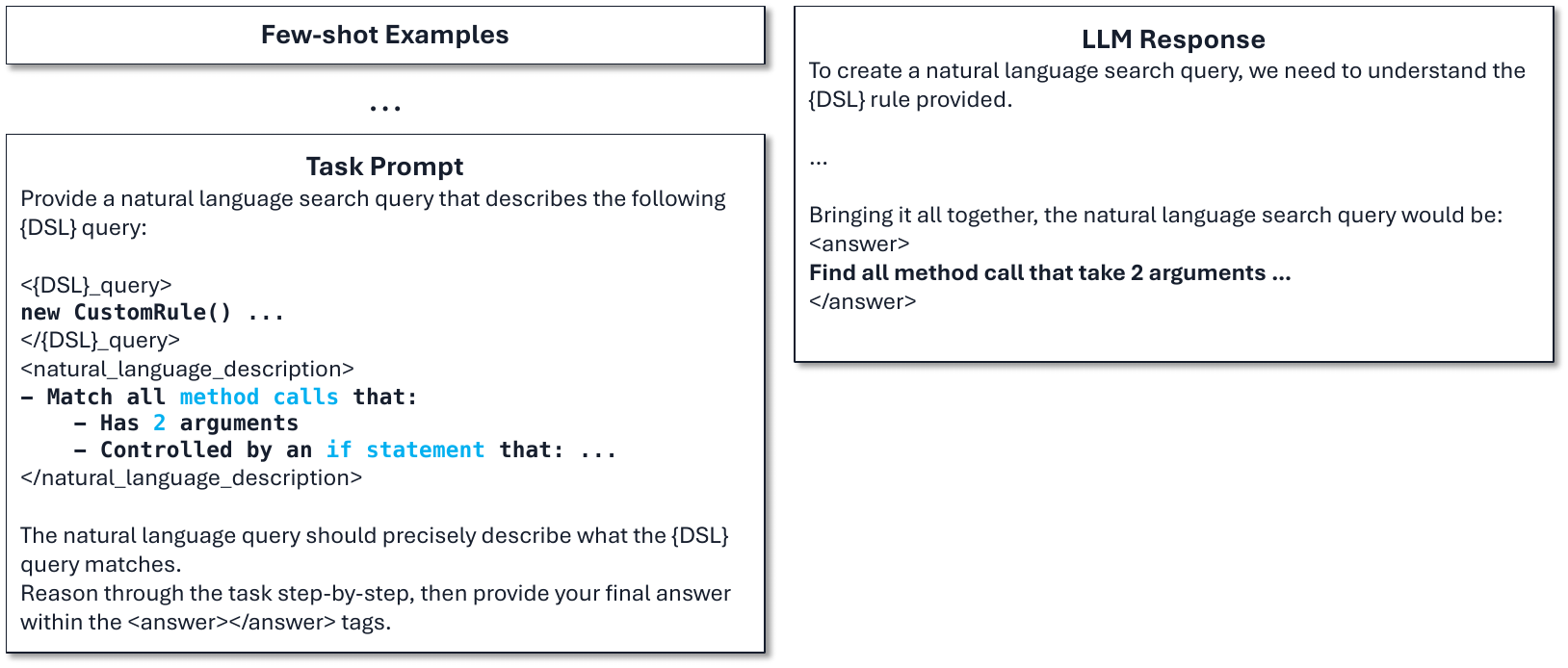}
    \caption{Prompt used to translate queries expressed in a DSL to natural language.}
    \label{fig:dsl-to-nl-prompt}
\end{figure}

Once a diverse set of queries is enumerated using the generation algorithm described in Section \ref{sec:query-enumeration}, each query is paired with its natural language equivalent by prompting an LLM to translate the DSL query to natural language.
As shown in Figure \ref{fig:dsl-to-nl-prompt}, the LLM is provided with both the query expressed in the DSL and an NL description in a structured format. 
To derive these NL description in structured format, each component of the DSL query (e.g., predicates in GQL and AST nodes in Semgrep patterns) is mapped to a NL description template.
We describe these templates for the GQL and Semgrep DSL in Section \ref{sec:implementation}.
%In Figure \ref{fig:dsl-to-nl-prompt}, each line is a template corresponding to a predicate 

To teach the LLM to perform this task accurately, we also provide the LLM with human-verified examples of the completed task (tuples of DSL query, NL description in structured format and target NL query in free-form text).
We construct these examples by initially prompting the LLM to complete the task in a zero-shot setting, followed by manually correcting the LLM's reasoning steps and the final answer.

Note that by generating DSL queries (as described in Section~\ref{sec:query-enumeration}) and then pairing these queries with equivalent NL queries, we obtain paired (NL, DSL) queries that are indexed in the RAG used by the LLM to translate NL query to DSL. Further, these (NL, DSL) pairs also serve as a benchmark dataset that we use for evaluation.

%% file: sections/approach/3-rag.tex
% \subsection{Retrieval Augmented Generation (RAG)}
% \label{sec:rag}

% Using a training dataset of NL and DSL structural code search query pairs, we compute an embedding for each NL query.
% When given an unseen NL query, our approach computes the embedding of the given query and retrieves 15 queries from the training set whose NL query embeddings have the highest cosine similarity. 

%% file: sections/approach/4-error-feedback.tex
\subsection{Query Refinement using Error Feedback}
\label{sec:error-feedback}

\begin{figure}[h]
    \centering
    \includegraphics[width=0.9\textwidth]{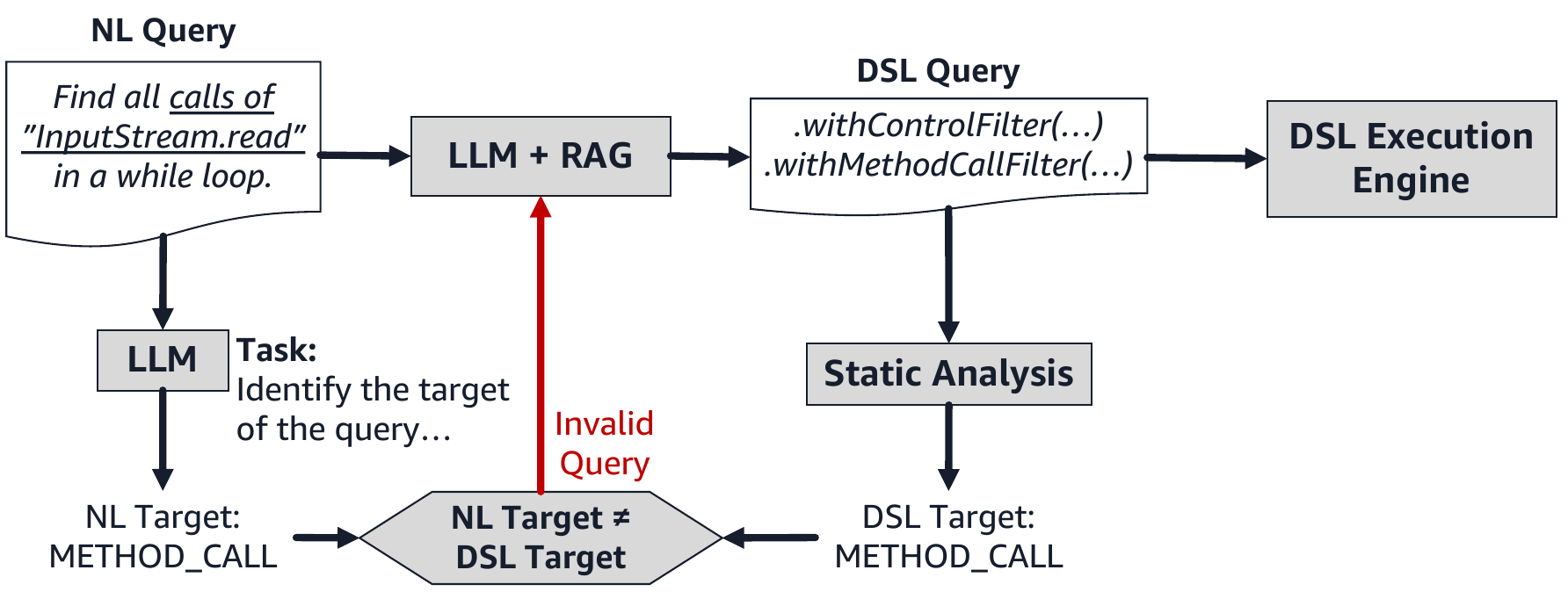}
    \caption{Our approach provides the LLM with feedback to refine the query using static analysis of the generated DSL query.}
    \label{fig:error-detection}
\end{figure}

As illustrated in Figure~\ref{fig:error-detection}, our approach improves the reliability of the NL to DSL translation by incorporating an error detection and feedback mechanism to refine DSL queries that may be incorrect.
%We provide two types of feedback to the LLM: errors reported by the structural code search engine and potential errors detected by statically analyzing the generated DSL query.

%\begin{lstlisting}[label={lst:example-error-detection}, caption={Example of an error returned by }]
%    
%\end{lstlisting}

A common error that the LLMs make when translating from NL to a DSL is to misidentify the target code construct of the query. For example, the NL query "\textit{Find all calls of \code{InputStream.read} in a \code{while} loop}" references two distinct constructs: a method call, and a while loop.
The target construct for this query is the method call \code{InputStream.read}.
However, while translating this query the LLM may incorrectly generate a DSL query that matches the while loop instead of the method call.
To verify that the DSL query matches the same code construct type as expressed in the NL query, we use the following approach. 
We separately prompt the LLM to identify the desired target code construct type in the given NL query. In addition, we statically analyze the DSL query generated by the LLM + RAG and determine the actual code construct type of this query.
If these two construct types do not match, we prompt the LLM with this feedback and re-generate the DSL query translation for the given NL query.
%as shown in Figure~\ref{fig:feedback-prompt}.

%% file: sections/12-implementation.tex
\section{Instantiating Structural Code Search Algorithm with DSLs}
\label{sec:implementation}

In this section, we describe how we instantiate our approach for structural code search with two different DSLs-- GQL and Semgrep. First, we describe the DSL-specific instantiations for the following functions from Algorithm~\ref{alg:dsl-generate}: \textsc{Init}, \textsc{Complexity}, \textsc{Specialize} and \textsc{Generalize}. We then describe the details for pairing DSL and NL queries for both the GQL and Semgrep DSL.

\begin{figure}
    \centering
    \begin{subfigure}{\textwidth}
    \centering
        \includegraphics[width=0.29\textwidth]{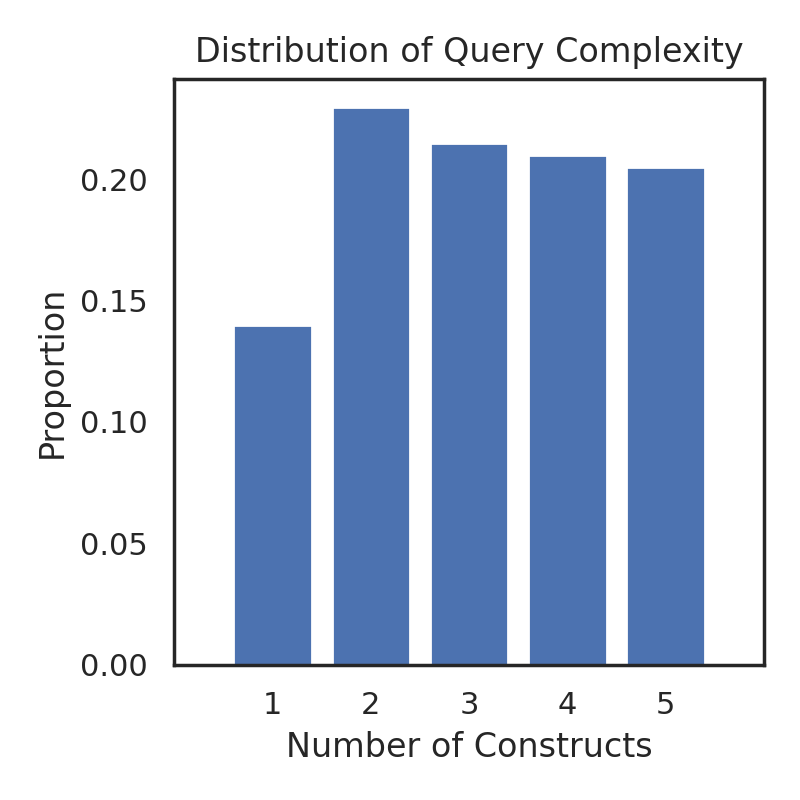}
        \includegraphics[width=0.49\textwidth]{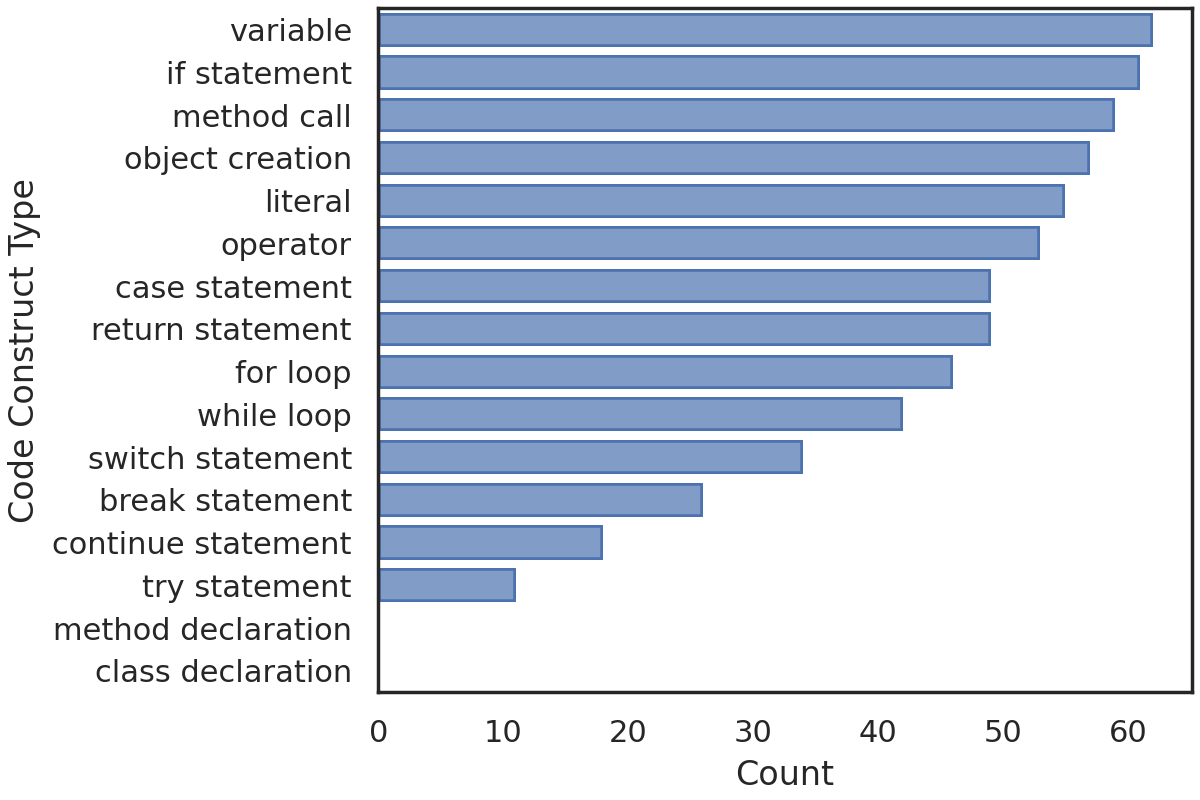}
        \caption{Distribution of number of code constructs and their types in the GQL-Full benchmark.}
    \end{subfigure}
    \begin{subfigure}{\textwidth}
    \centering
        \includegraphics[width=0.29\textwidth]{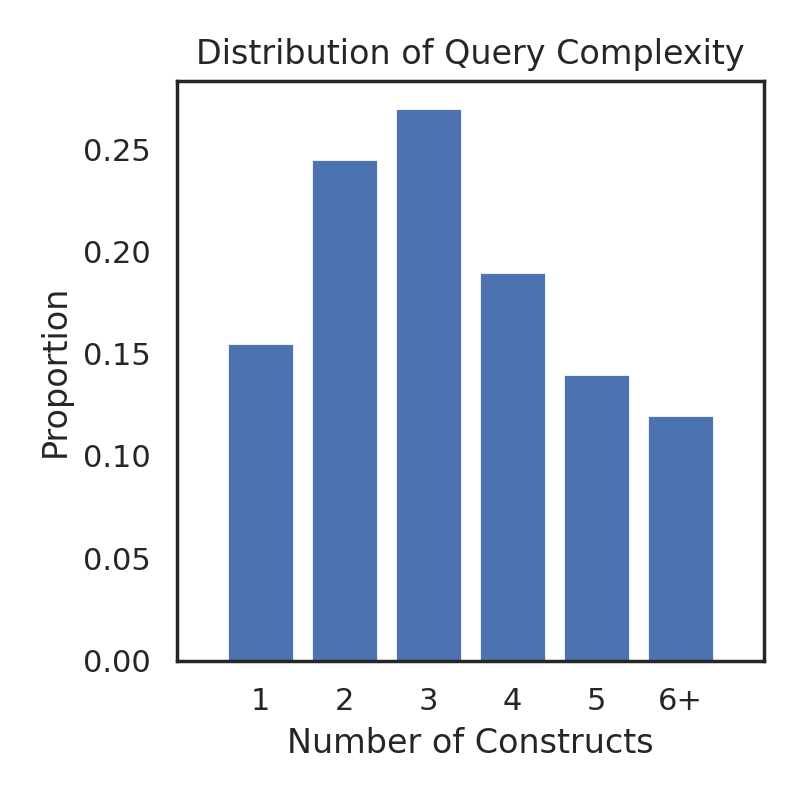}
        \includegraphics[width=0.49\textwidth]{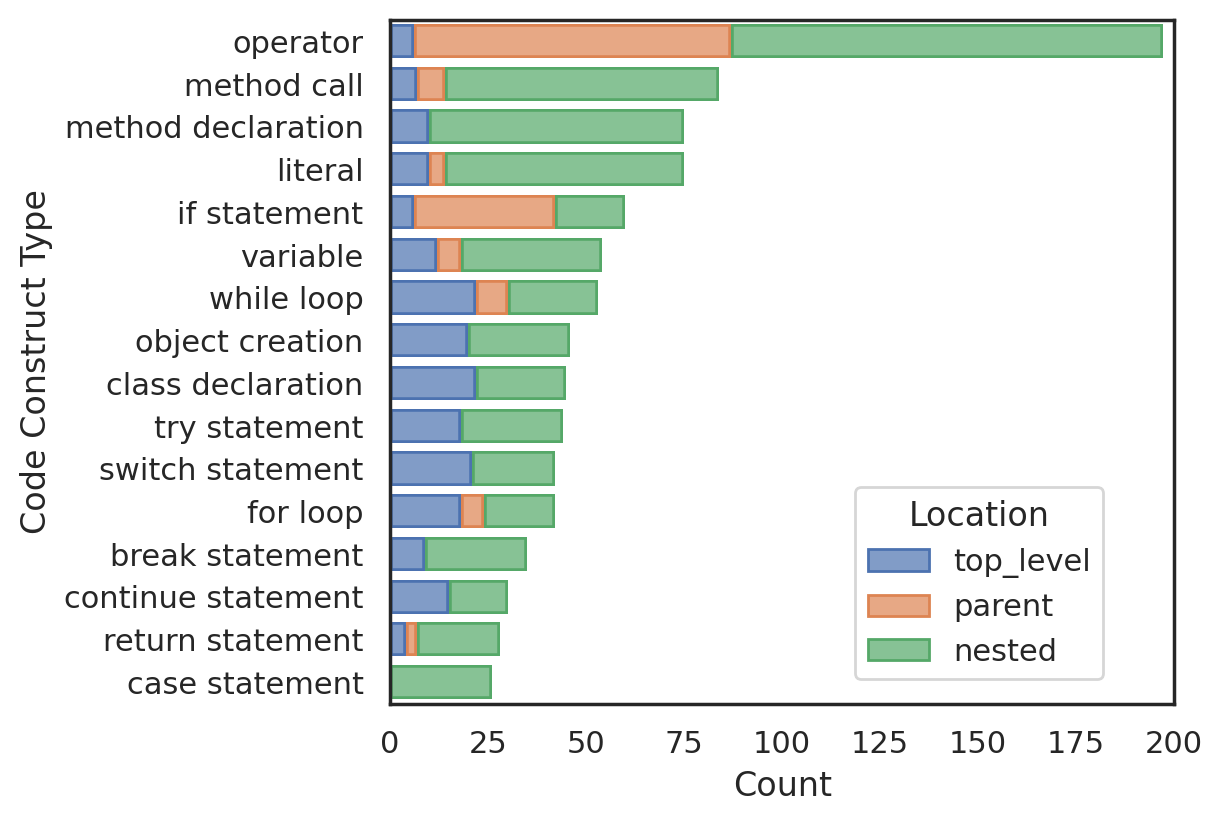}       
        \caption{Distribution of number of code constructs and their types in the Semgrep-Full benchmark.}
    \end{subfigure}
    \caption{Distributions of query complexity and the occurrence of code constructs in the GQL-Full and Semgrep-Full benchmarks. Since GQL does not support matching method declarations and class declarations counts of these constructs is zero.}
    \label{fig:benchmark-stats}
\end{figure}

\subsection{Instantiating structural code search algorithm with GQL}

\noindent\textit{GQL query enumeration:} \\
GQL is an imperative language with a query constructed by chaining together different predicates. Algorithm~\ref{alg:dsl-generate} for GQL enumerates queries by starting with the most general query (that has a single predicate that canonically targets the chosen code construct type) and specializing it-- one predicate at a time, till the target complexity is achieved. 
% A query $q$ in GQL is thus a sequence or chain of GQL predicates, denoted as $q=(q_1, q_2, \dots, q_n)$ where each $q_i$ for $i\in[1, n]$ is a GQL predicate. 

\textsc{Init}$(t)$: 
Given a target code construct $t$, \textsc{Init} returns a GQL query with a single GQL filter operation that corresponds to the type of $t$. For example, if $t$ is the \code{for} statement in Listing \ref{lst:impl-example-code}, the GQL query \textsc{Init} returns the query $q$ comprising the  \code{withControlFilter("FOR\_STATEMENT")} operation. The \textsc{Init} query is constructed by consulting a mapping from code construct types and their corresponding GQL filters. 

% \textsc{Execute}$(q, T)$:
% The \textsc{Execute} function returns the code construct instances that result from compiling and executing the query $q$ over a code corpus $T$ using a standalone GQL command line tool. The output of the tool is post-processed to extract the line number of each code construct matched by the query.

\textsc{Complexity}$(q)$: The complexity is computed as the number of GQL predicate groups each comprised of a transform and a set of successive filter operations that match a code construct. 
%\pranav{is this correct?} \ben{yes}
% chained GQL predicate groups (a transform and a set of filter operations that match some code construct) in the query $q$.

\begin{lstlisting}[language=Java, label={lst:impl-example-code}, caption={An example Java program to be matched by a generated query.}]
String a;
for (int i = 0; i < limit; i++) {
a += Integer.toString(number);
}
\end{lstlisting}
\begin{lstlisting}[language=Java, label={lst:gql-rule-generated}, caption={A generated GQL query.}]
new CustomRule.Builder()
    .withName("Generated Query")
    .check()
    // Added by the initialization step:
    .withControlFilter("FOR_STATEMENT") 
    // Added by the specialization step:
    .withContextNodesTransform(ContextKind.LOOP) 
    .withNodeByTypeFilter(EGroumASTNodeType.METHOD_INVOCATION)
    .withMethodCallFilter("java\\.lang\\.Integer\\.toString")
    .build()
\end{lstlisting}

\textsc{Specialize}$(q, t, T, Q)$: The \textsc{Specialize} function for GQL is implemented by speculatively executing transforms to identify a code construct $t'$ that is related to the current target code construct $t$ by some relation.
In other words, given a query $q$, \textsc{Specialize} speculatively calls $\textsc{Execute}(q \oplus r, T)$ where $r$ is a transform that relates $t$ with another construct in the larger code context $T$.
This results in a set of possible new target constructs $T_{next}=\bigcup_{r\in R} \textsc{Execute}(q \oplus r, T)$ where $R$ is the set of available GQL transforms at the current code construct $t$.
 As an example, in Listing \ref{lst:impl-example-code}, \textsc{Specialize} algorithm-- when called with $t$ being the \code{for} statement, may execute the \code{withContextNodesTransform(...)} transform and identify the \code{+=} operator, \code{toString} method call and variables \code{a} and \code{number} on line 3 as potential new target constructs $t'$.

To bias towards a uniform distribution of code construct types in $Q$, the new target construct $t'$ returned by \textsc{Specialize} is selected from $T_{next}$ with the probability $1/(1+c_e)$ where $c_e$ is the number of instances of the construct type $e$ in $Q$.

\begin{figure}[h]
    \centering
    \includegraphics[width=0.9\textwidth]{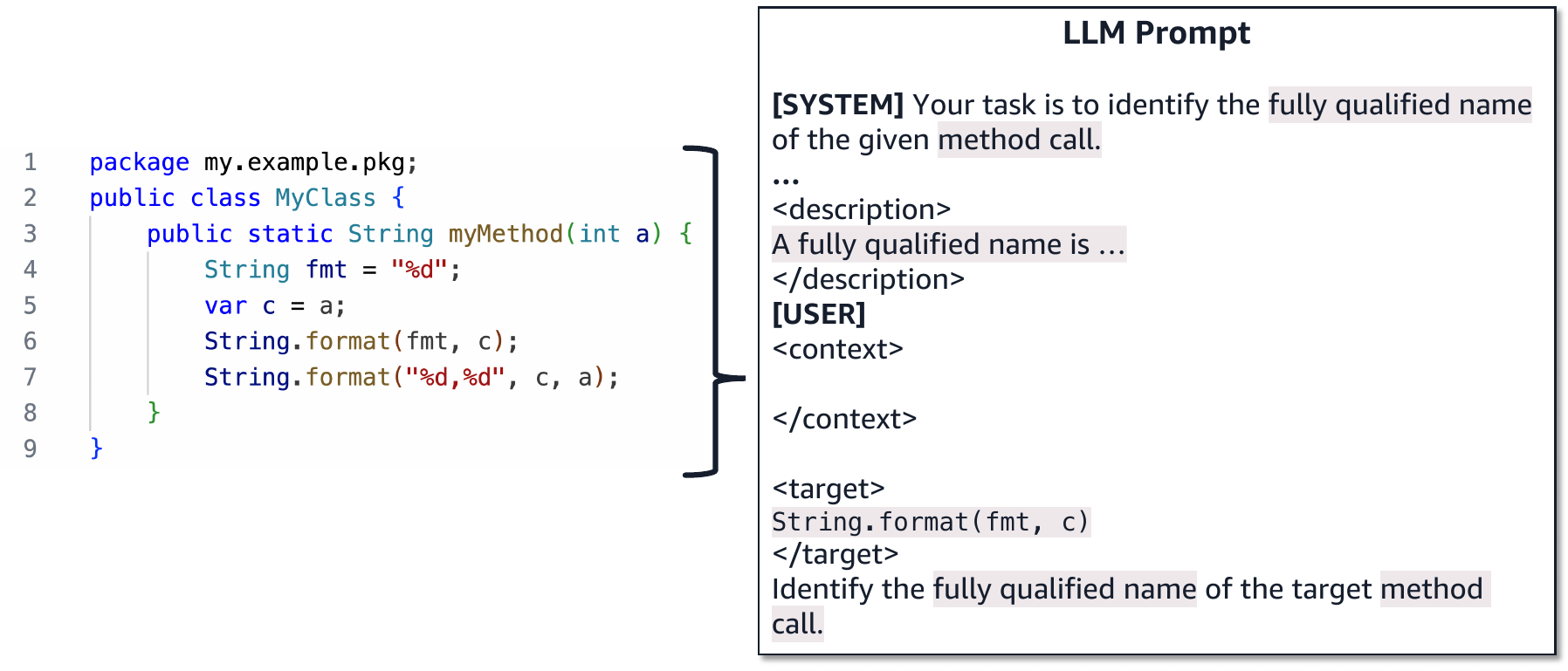}
    \caption{The LLM prompt used to extract properties of the target code construct to instantiate in a GQL predicate.}
    \label{fig:gql-llm-gen-prompt}
\end{figure}

Lastly, to finalize the specialization step, any missing arguments required by the selected transform and filter operations are instantiated by querying an LLM to identify the properties of relation between $t$ and $t'$ or the properties of  the target construct $t'$.
Continuing with the above example, if the \code{toString} method call on line 3 of Listing~\ref{lst:impl-example-code} was selected as the new target construct $t'$, then a GQL filter operation may be instantiated with its fully qualified name using an LLM prompt as shown in Figure~\ref{fig:gql-llm-gen-prompt} .

\textsc{Generalize}$(q, Q)$: 
Since GQL query enumeration algorithm starts with the most general query (with no predicate) and incrementally specializes it, the \textsc{Generalize} function is never called (the complexity of query $q$ is never  greater than $c_{max}$). \\

\begin{lstlisting}[label={lst:gql-template}, caption={Example of a declaratively defined template for a GQL filter operation that conditions on the number of arguments of a method call.}]
GQLValuePropBuilder(
    name="argument count",
    gql_template=".withNumberOfArgumentsFilter({value})",
    natural_template="has {value} arguments",
    description="For example, the method call `MyClass.foo(abc, def)` has 2 arguments.", )
\end{lstlisting}

\noindent \textit{Pairing GQL and NL queries}: 
As described in Section~\ref{sec:deriving-nl-dsl-pairs}, to translate a GQL query into NL query we provide a programmatic translation from GQL to an NL query description in a structured format. This structured format is in the form of a template that describes all the GQL filter or transform operations. 
% To provide the natural language query description used to prompt an LLM to translate the generated GQL query to natural language (see Section \ref{sec:deriving-nl-dsl-pairs}), 
% we associate a natural language template that describes the GQL filter or transform that can be used in a query.
This is a declarative format and support for new filter or transform operations can be added in a few (<5) lines of code (refer to Listing~\ref{lst:gql-template}).

As an example, \code{natural\_template} in the  \code{GQLValuePropBuilder} in Listing~\ref{lst:gql-template} is instantiated with the parameter value passed to the \code{withNumberOfArgumentsFilter} operation in the given GQL query. This instantiated string \textit{"Has 2 arguments"} becomes part of the structured NL description such as the one included in the prompt in Figure~\ref{fig:dsl-to-nl-prompt}.

\subsection{Instantiating structural code search algorithm with Semgrep}

\noindent \textit{Semgrep query enumeration:}

\textsc{Init}$(t)$: For the target code construct $t$, \textsc{Init} returns the Semgrep query with a \code{pattern} that matches $t$. 
As an example, if the target construct $t$ were the add operator on line 3 in Listing \ref{lst:impl-example-code}, \textsc{Init} would return the following Semgrep query: \code{pattern: a += Integer.toString(number);} (also shown in Listing \ref{lst:semgrep-generated}).

% \textsc{Execute}:
% The \textsc{Execute} function returns the code construct instances that result from compiling and executing the query $q$ over a code corpus $T$ using Semgrep's standalone command line tool.
\textsc{Complexity}$(q)$: 
Returns the number of code constructs (e.g., operators, literals, meta-variables, ellipses, etc.) in the query $q$.

\vspace{2em}

\begin{lstlisting}[language=Java, label={lst:semgrep-generated}, caption={Steps of Semgrep rule generation}]
// -------Initial Rule--------
patterns:
  pattern: |
    a += Integer.toString(number);

// -------Specialized Rule--------
patterns:
  pattern: |
    a += Integer.toString(number);
  pattern-inside |
    for (int i = 0; i < limit; i++) {
    ...
    }
// -------Generalized Rule--------
patterns:
  pattern: |
    a += Integer.toString($METAVAR0);
  pattern-inside |
    for (int i = 0; ...; i++) {
    ...
    }   
\end{lstlisting}

\textsc{Specialize}$(q, t, T, Q)$: 
The \textsc{Specialize} function returns a query that conjoins the given query $q$ with a \code{pattern-inside} clause. 
%which enforces the constraint that the initial pattern must match within the pattern described in the \texttt{pattern-inside} clause.
We generate an appropriate pattern for the \code{pattern-inside} clause by identifying an enclosing context of the current code construct $t$.
We do so by by first parsing the source code file of $t$ into an abstract syntax tree (AST). 
%uniformly choosing a parent node of $t$ in the abstract syntax tree of the target code
We then uniformly select an ancestor node of $t$ in the AST and use the code that corresponds to the sub-tree at the ancestor as the field in the \code{pattern-inside} clause. %\pranav{parent node of t or an ancestor of t?} 
As an example, the original Semgrep rule with the add operation may be \textsc{Specialize}d by adding a \code{pattern-inside} clause with the \code{for} loop construct as shown in Listing \ref{lst:semgrep-generated}. 
%As explained, this would happen by selecting the AST node corresponding to the \code{for} statement at line 2 that is a parent node of the initial add operation in the AST.
At the same time, we also replace construct $t$ in the sub-tree at the ancestor node with an ellipsis. 
This allows the \code{pattern} clause of $q$ to be  independent of the \code{pattern-inside} clause with which it is conjoined.
%of the corresponding to construct $t$ to be independently generalized, we replace the AST node corresponding to the code construct $t$ with an ellipsis in the \code{pattern-inside} clause.
As the \textsc{Specialize} function does not change the target code construct, it returns $t' \gets t$.

\textsc{Generalize}$(q, Q)$: 
The \textsc{Generalize} function first parses all the code patterns inside the \code{pattern} or \code{pattern-inside} clause of the query $q$ into an AST. It then selects a node in these ASTs and returns an updated query $q'$ that replaces the selected node with an ellipses or a meta-variable. 
As an example, the generalization in Listing~\ref{lst:semgrep-generated}  replaces the AST node that corresponds to the variable \code{number} with a meta-variable \code{\$METAVAR0} (line 17), and replaces the expression \code{i < limit} inside the \code{pattern-inside} clause with an ellipses (line 19).
Generalizing concrete AST nodes in the pattern with meta-variables or ellipses reduces the complexity of the query by removing one more code constructs.

We choose the AST node to replace with a probability weighted towards AST nodes that contain more frequently sampled code construct types. This increases the likelihood of the constructs already present more frequently in $Q$ to be generalized. 
Concretely, for an AST node $a \in q$, $w_a = \sum_{c\in a} \textsc{CountType}(c, Q)$ where $c$ is an AST node in the sub-tree at $a$ and \textsc{CountType} returns the frequency of the construct type of $c$ in the set of previously generated queries $Q$. Then, 
we choose the AST node $a$ for to be generalized with probability $w_a/\sum_{b\in q} w_b$.\\

\noindent \textit{Pairing Semgrep and NL queries}: 
As described in Section~\ref{sec:deriving-nl-dsl-pairs}, to translate a Semgrep query into NL query we provide a programmatic translation from Semgrep to an NL query description in a structured format. This structured format is 
a serialization of the AST of the Semgrep \code{pattern} and \code{pattern-inside} clauses as a nested list. Refer to Listing~\ref{lst:semgrep-nl-desc} for an example of the NL description in structured format for an example Semgrep query. 

\begin{lstlisting}[label={lst:semgrep-nl-desc}, caption={An example of a pattern description that would be provided to the LLM. The description is derived from the AST of the Semgrep pattern.}]
<semgrep_pattern>
    while (! $VAR1 .interrupted()) { ... }
</semgrep_pattern>
<pattern_description>
- while_statement
  - condition: parenthesized_expression
    - unary_expression
      - operand: method_invocation
        - object: metavariable: '$VAR1'
        - name: identifier: 'interrupted'
        - arguments: argument_list
  - body: block
    - ...
</pattern_description>
\end{lstlisting}

%% file: sections/13-evaluation.tex
\section{Evaluation Setup}
\label{sec:eval-setup}
\subsection{GQL-Derived and Semgrep-Derived Benchmarks}

Using the algorithm described in Section \ref{sec:approach}, we construct two structural code search benchmarks: one derived from GQL queries and one derived from Semgrep queries. Each benchmark consists of 200 structural code search queries. Each query consists of a natural language query and its corresponding representation in the respective DSL (GQL or Semgrep). Each DSL query is executed using its corresponding static analysis engine over a code corpus of 10 Java projects from IJaDataset 2.0 \cite{bigclonebench-ijadataset}. In total, these 10 projects contain 702 source files with 76,446 lines of code (or ~4.3 megabytes of code). 
%\pranav{why these 10 packages?}\ben{added next line to explain}
The projects were selected on the basis of their license (MIT or Apache-2.0) and size (between 90-200 classes per project).
Since some projects may use more restrictively licensed code, we only include source files that contain an explicit license header.
For each query, the corresponding matched lines of code reported by GQL or Semgrep are recorded.

For each full GQL-derived and Semgrep-derived benchmark, we also designate a lite version of each benchmark consisting of a randomly selected subset of 10 queries and 100 source files with 27,252 lines of code or 1.3 megabytes of text. %(\pranav{say these are to benchmark a purely LLM baseline that is both slow and expensive}).
The purpose of the lite subsets is to benchmark pure-LLM baselines which are slow and expensive in terms of tokens to execute over a large corpus. 

To mitigate contamination between the benchmarks and the training sets used for RAG in our approach, we ensure that both the code corpuses and queries are disjoint between datasets.
%\pranav{say examples in the benchmark are disjoint from the ones used as few-shot examples in RAG}\ben{added line}

\subsection{Baselines}
%By leveraging the same algorithm described in Section \ref{sec:approach}, we generate a training dataset of (NL, DSL) query pairs to be used as few-shot examples for NL-to-DSL translation. 

%Both the NL-to-GQL and NL-to-Semgrep baselines use GPT-4o-2024-08-06 unless otherwise stated (\pranav{this should be moved to a separate implementation section; this is not a baseline}). Both baselines translate each natural language structural code search query to GQL with few-shot examples retrieved using vector search with bge-small-en-v1.5 as the embedding model (\pranav{embedding should feature in the Approach overview figure}). Both baselines also use an automatic re-prompting strategy to verify and repair the generated DSL queries (\pranav{explain this in Sec 3}). The key difference between the two baselines is the choice of the code search backend (either GQL or Semgrep).

We compare our approach against two different baseline approaches based on LLMs and vector similarity search.

\begin{figure}[h]
    \centering
    \includegraphics[width=0.9\textwidth]{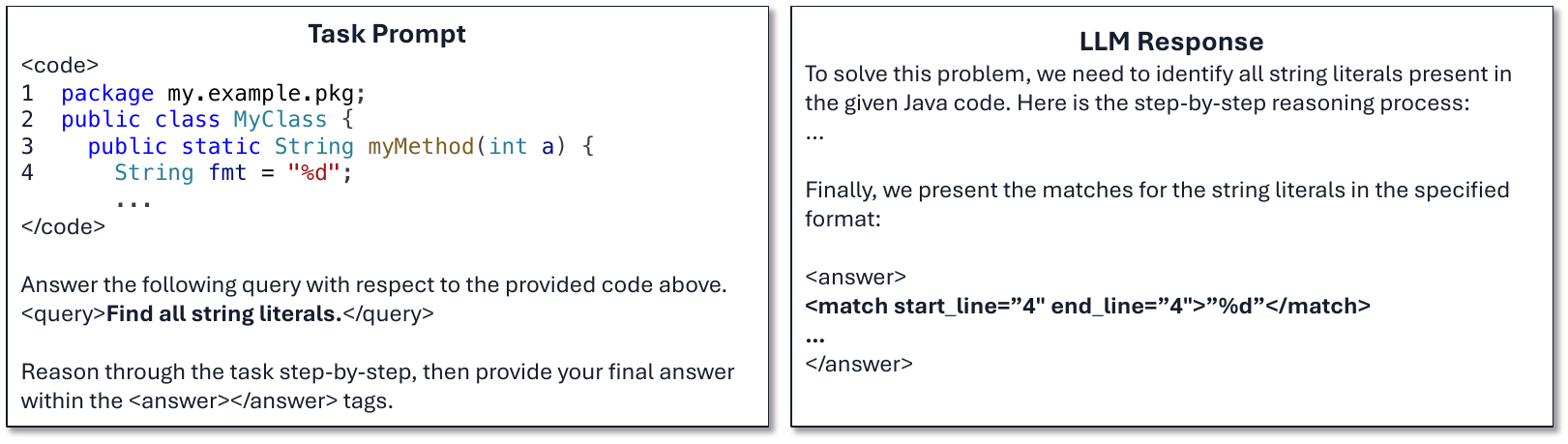}
    \caption{Example prompt used for In-Context Retrieval baselines.}
    \label{fig:pure-llm-prompt}
\end{figure}

We implement the pure LLM baseline by prompting an LLM to answer the NL search queries directly. For each source file in the code corpus, the LLM is prompted with the content of the source file and asked to answer the query by with the relevant lines of code. Figure \ref{fig:pure-llm-prompt} shows an example of the prompt and expected response.
%\pranav{there is no in-context retrieval in this approach IMO}\ben{should we rename this baseline? e.g. Pure LLM?}

Since the pure LLM baselines rely on prompting with the code to be searched over per file, the full 400 query benchmark over a 10 project database consisting of 76k lines of code is too large to practically evaluate the baseline.
As such, we only compare the performance of our approach against pure LLM baselines on the Lite benchmarks.

Vector similarity search techniques retrieve relevant code based on the similarity of its embedding with the embedding of the query.
Vector search offers greater speed and efficiency, but can only perform retrieval at the granularity of the chunk size chosen.
Therefore, to compare vector search against other baselines, we chunk code at the method-level to provide sufficient context within each chunk to answer most queries.
%\pranav{does method level chunking conflict with matching code using just the first line of the chunk?}
%\ben{not sure I understand how it would?}

For our evaluation, we choose NV-Embed-v2: the current top-performing embedding model on the Massive Text Embedding Benchmark (MTEB) as of November 2024.
For each query, we compute the cosine similarity $s$ of the embedding vector of the query $E_Q$ with the embedding of each method $m$ in the code corpus $E_{m}$ as $s = E_Q \cdot E_m / (\|E_Q\|\|E_m\|)$.
We then retrieve all methods whose cosine similarity passes a given threshold $T$ which we vary in our evaluation.

\subsection{Metrics}
To evaluate the performance of our NL-to-DSL translation approaches and to compare them against other baselines, we measure the recall, precision, and F1 score averaged across all queries.

Each query in our benchmark dataset can have one or more \emph{code matches}. Each code match consists of a span of code matched in the code corpus. We use the start line of code span to match code.

\begin{itemize}
    \item The \textbf{recall} of one query in the benchmark is computed as $TP_G/P$ where $TP_G$ is the number of ground-truth code matches that have an equivalent predicted code match, and $P$ is the total number of ground-truth code matches.
    \item The \textbf{precision} of one query in the benchmark is computed as $TP_P/PP$ where $TP_P$ is the number of predicted code matches that have an equivalent ground-truth code match, and $PP$ is the total number of predicted code matches.
    \item The \textbf{F1 score} is the harmonic mean of the recall and precision: $2 * \frac{recall \times precision}{recall + precision}$
\end{itemize}

The overall recall, precision, and F1 scores is the average of the recall, precision, and F1 scores computed for each query in the benchmark.

\section{Evaluation Results}
\label{sec:evaluation}

In this section, we answer the following research questions:

%\ben{
%- application of this work, justify it use based on prior work
%- approach - compare pure LLM, pure RAG, NL-to-DSL, 
%- then comparison between semgrep and GQL
%
%- in approach/implementation - highlight the \# of lines of code for adding support for a new DSL
%
%High level:
%How well does this perform for the application
%
%How does our approach compare to other approaches
%
%What are the insights of our approach for various DSLs
%- break down the NL-to-DSL translation
%- insights on where the NL-to-Semgrep and NL-to-GQL differ
%- limitations of the tools - what constructs/formats does it not work well in
%
%
%Move RQ4 to the top, combine RQ1 and 3, Move RQ2 after RQ4.
%
%
%}

\begin{enumerate}[label=\textbf{RQ\arabic*:}]
%    \item \label{rq-survey} \ben{ETA: 1 day} What is the space of natural language queries that are expressable across other strucutral code search DSLs?
%    \ben{i.e. The limitations of this work. Qualitative literature survey of queries from other code search tools.}
% move to background
%    \item \label{rq-nl-to-dsl} \ben{ETA: 8 days} How do different approaches to answering structural code search queries compare in terms of accuracy and efficiency?
%    \ben{We'll measure accuracy in terms of recall/precision/f1 score using our benchmark; Efficiency is measured in number of tokens and end-to-end speed. This is also where we'll compare against the pure LLM and pure RAG baselines. We should also include a qualitative analysis of the results to discuss the quirks and limitations of each approach.}
    \item \label{rq-performance} How effective is NL-to-DSL approach at answering natural language structural code search queries?
    %\item \label{rq-comparison} How do NL-to-DSL translation baselines compare with other baselines?
    %\ben{Ablation study of our LLM+RAG+Semgrep/GQL solutions. Also compare against just using the GQL/Semgrep documentation in RAG or pre-existing rule examples}
    \item \label{rq-comparison} How performant is NL-to-DSL translation approach for structural code search compared to baselines?
    \item \label{rq-ablation} How does the performance depend on the choice of the RAG index and the query refinement module?
    
    %\ben{Measured by comparing code search tools using a semgrep or GQL backend on the semgrep and GQL derived benchmarks.}
    %\item \label{rq-correctness} How accurately do the synthetic natural language queries describe the structural code search queries?
%    \ben{We sample 10\% of the generated queries and manually verify that the semantics of the natural language query match the DSL query.}
    %\item \label{rq-robustness} \ben{ETA: 3-5 days?} How robust is the natural language query translation?
    %\ben{We'll use semantic-preserving text mutations to perturb the NL queries and measure how it affects the performance of the code search tools}
\end{enumerate}

We follow these with a discussion on the limitations of our proposed approach and threats to validity of the evaluation results.

\input{sections/RQs/rq-performance}
\input{sections/RQs/rq-ablation}
\input{sections/RQs/rq-breakdown}

%% file: sections/RQs/rq-performance.tex
\subsection{\ref{rq-performance} Effectiveness of NL-to-DSL translation approach at structural code search}
%\pranav{focus on the performance of NL-to-GQL on the GQL dataset and NL-to-Semgrep on the Semgrep dataset. Showcase that the approach works. Provide a few intuitive examples to drive home the utility of this capability}

\begin{table}[h]
\centering
\caption{Performance of NL-to-DSL translation approach for structural code search.}
\label{tbl:performance-table}
\begin{tabular}{lclrrr} 
\toprule Benchmark& Granularity of code match & Rec. (\%) & Prec. (\%) & F1 (\%)  \\ 
\midrule
 GQL-Full     & Line& 59.9& 57.3&58.5\\
 Semgrep-Full & Line& 70.6& 69.4&70.0\\
\bottomrule
\end{tabular}
\end{table}

\begin{figure}[h]
    \centering
    \includegraphics[width=0.49\textwidth]{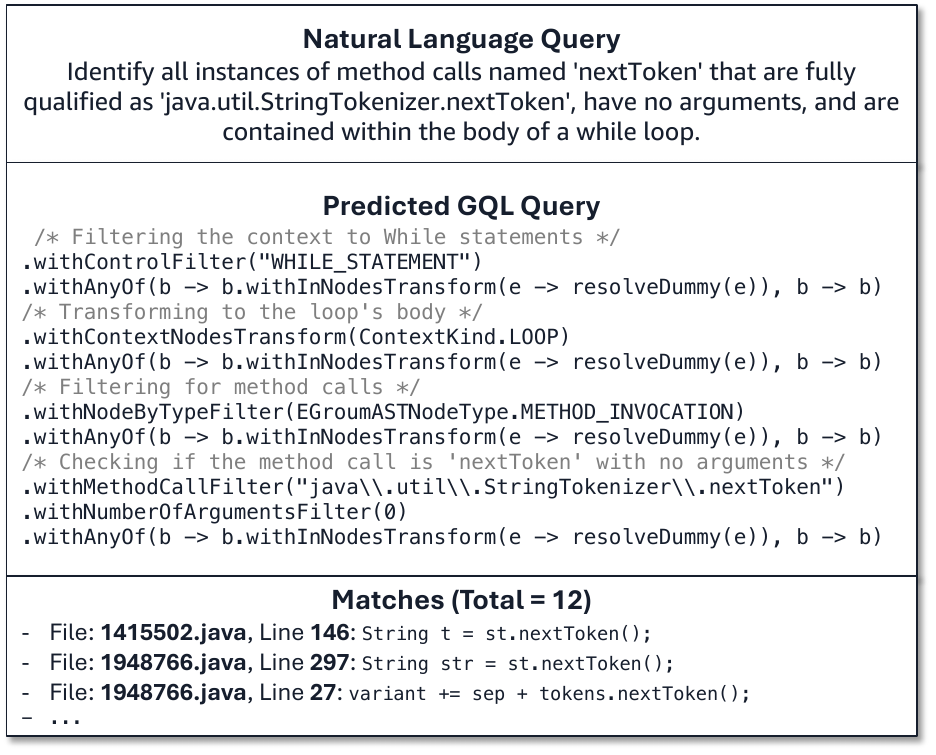}
    \includegraphics[width=0.49\textwidth]{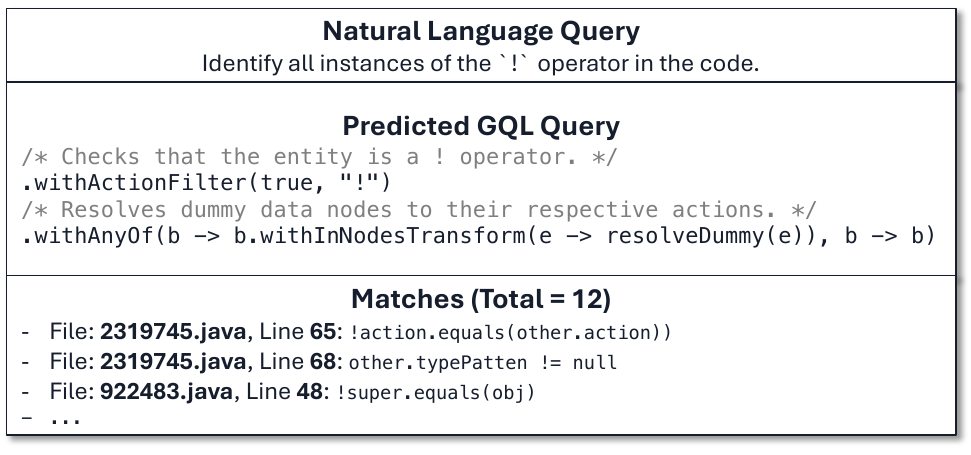}
    %Placeholder
    \caption{Example of a complex and a simple NL query that is accurately matched to code by our structural code search approach.}
    \label{fig:gql-query-examples}
\end{figure}

\begin{figure}[h]
    \centering
    \includegraphics[width=0.49\textwidth]{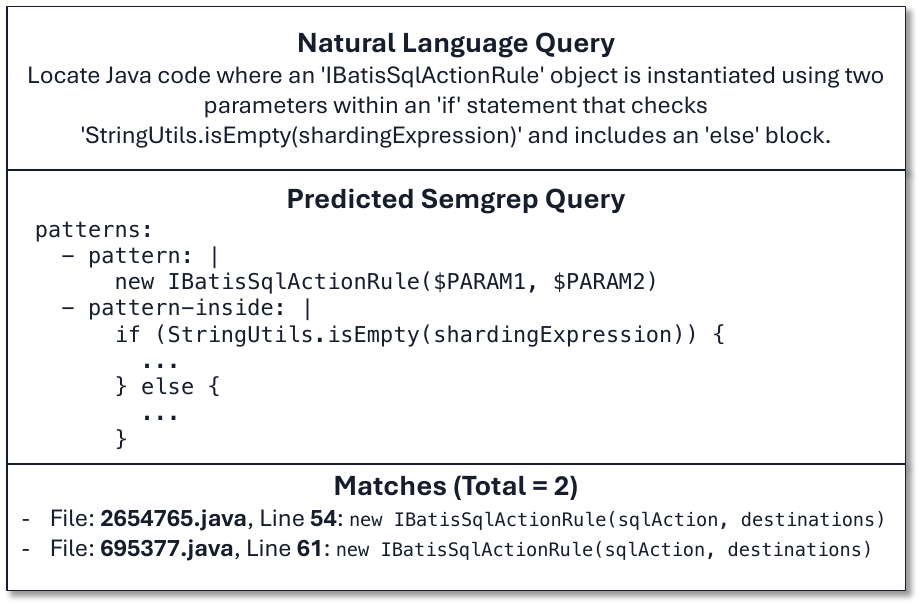}
    \includegraphics[width=0.49\textwidth]{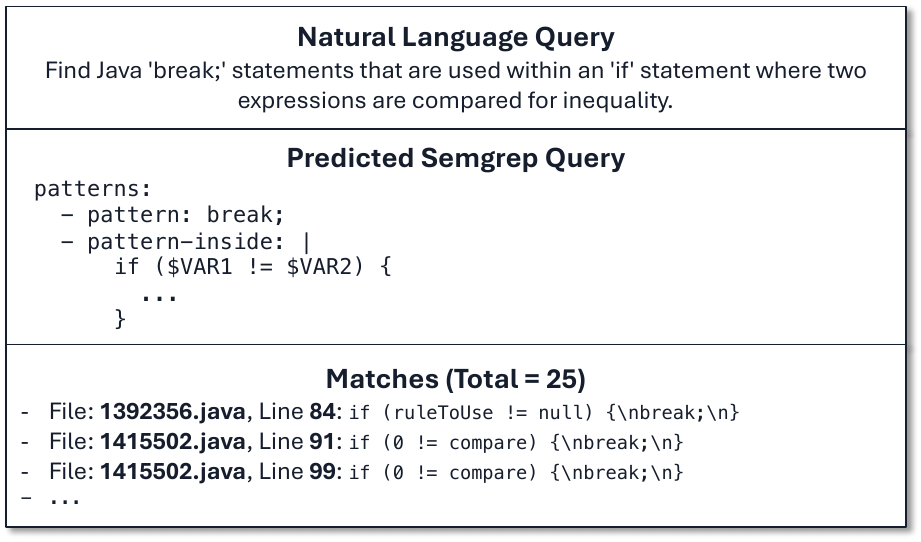}
    \caption{Example of a complex and a simple NL query that is accurately matched to code by the NL-to-Semgrep translation approach.}
    \label{fig:semgrep-query-examples}
\end{figure}

We evaluate our NL-to-DSL translation approach for structural code search on both the GQL and Semgrep derived benchmark datasets (Table~\ref{tbl:performance-table}). The precision, recall and F1 scores all lie within a high range of 55\% - 70\%. 
In Figure~\ref{fig:gql-query-examples} and Figure~\ref{fig:semgrep-query-examples}, we illustrate few natural languages queries that are accurately matched to code using our translation approach. As we can see, these queries cover a variety of different code constructs -- for e.g., number of parameters in a method call, checking for conditionals with inequality checks, searching for a given operator, or checking on given fully qualified types for method calls. 
Furthermore, the search constraints that we handle can vary in terms of their complexity from comprising a single code construct to conjunctions of several constructs. 

%Give some examples of the variety of queries it successfully matches emphasizing the utility of this solution. 

\begin{figure}
    \centering
    \includegraphics[width=0.9\textwidth]{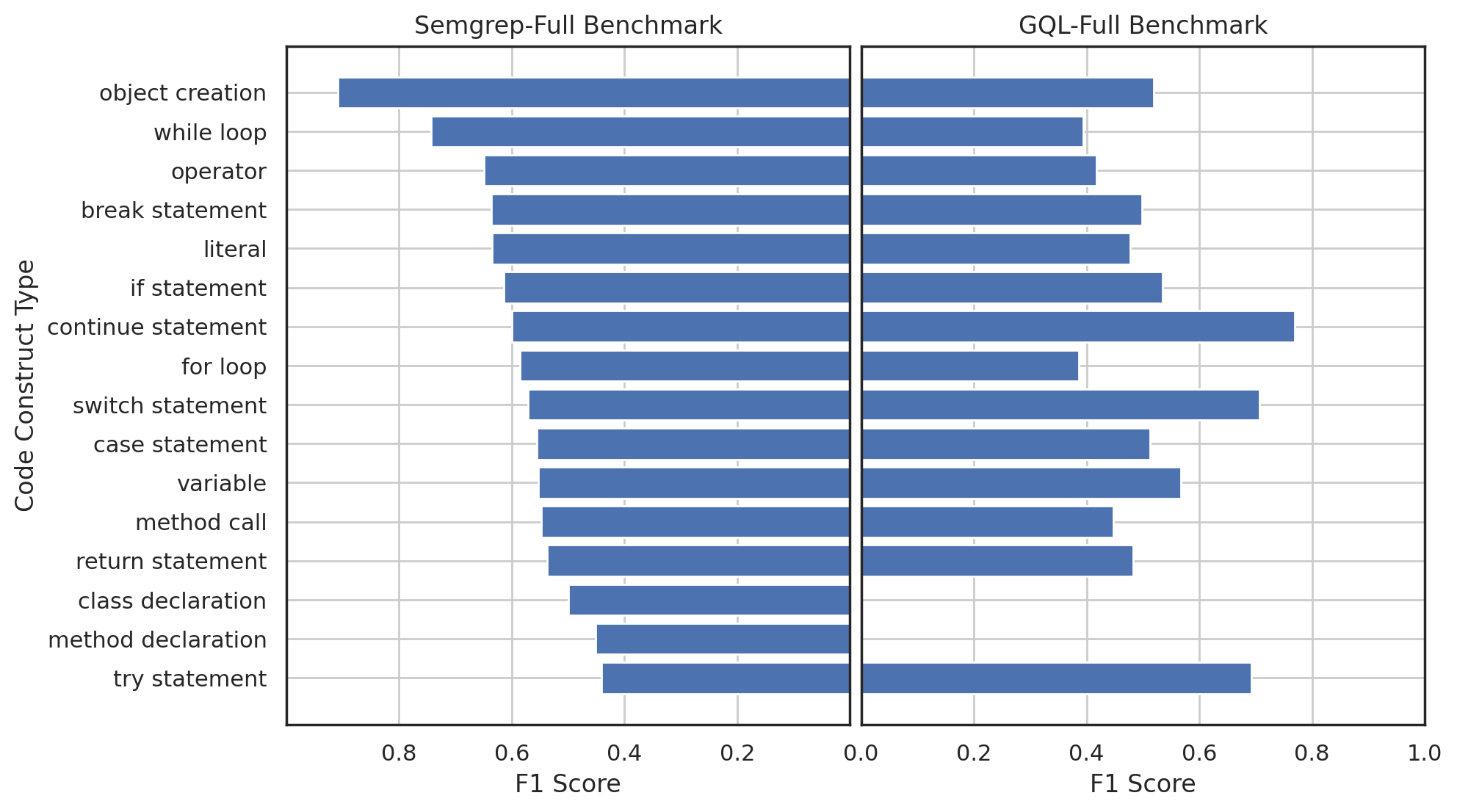}
    \caption{Performance on search queries comprising different code entities (NL-to-Semgrep on the left and NL-to-GQL on the right)). Note, GQL does not support queries that match class or method declarations.}
    \label{fig:f1-by-construct-type}
\end{figure}

\begin{figure}[ht]
    \centering
    \begin{subfigure}{0.4\textwidth}
        \includegraphics[width=\textwidth]{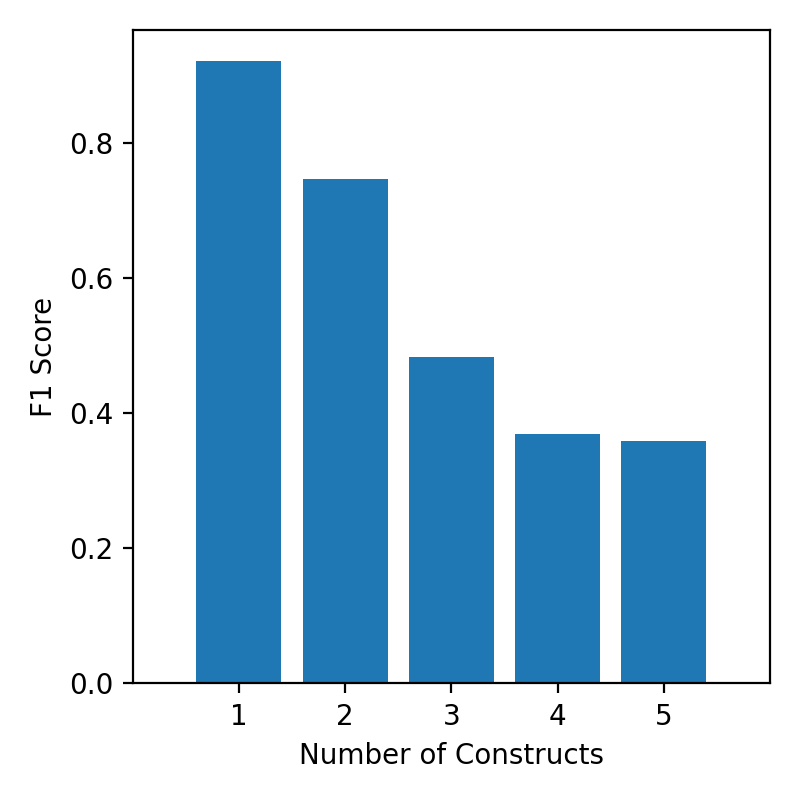}
        \caption{GQL-Full Benchmark}
    \end{subfigure}
    \begin{subfigure}{0.4\textwidth}
        \includegraphics[width=\textwidth]{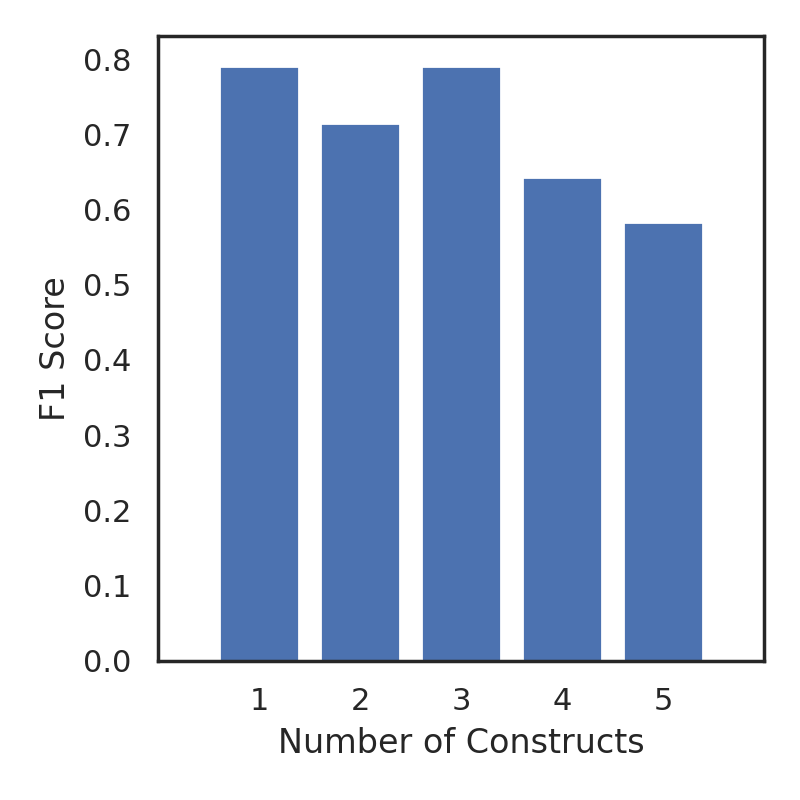}
        \caption{Semgrep-Full Benchmark}
    \end{subfigure}
    \caption{Performance on queries comprising up to 5 DSL constructs.
    %\pranav{should we remove the 6+ bar in the semgrep sub-figure}\ben{added a 5-only version commented above}
    }
    \label{fig:f1-by-num-constructs}
\end{figure}

To better understand the performance of NL-to-DSL on search queries involving different code constructs and with varying complexities, we break down the performance along these two axes in Figure~\ref{fig:f1-by-construct-type} and Figure~\ref{fig:f1-by-num-constructs}. While the performance does vary across queries with different code entities, we observe that our NL-to-DSL approach is robust and the performance holds up with an F1 score greater than 38\% for all query segments comprising different code construct types (Figure~\ref{fig:f1-by-construct-type}). 
Along the same lines, performance holds up as the queries become more complex with F1 scores greater than 35\% for GQL queries with up to 5 constructs and 58\% for Semgrep queries with up to 5 constructs (Figure~\ref{fig:f1-by-num-constructs}).
We vary the query complexity in our analysis till up to 5 constructs since several surveys of real-world usage indicate that developers often search with short queries with up to 5 terms per query~\cite{Bajracharya-queries-are-short, Sim-queries-are-short, pradel-code-search-survey}.

\begin{tcolorbox}
The NL-to-DSL translation approach is very effective at structural code search with a high precision and recall ranging from 55\% - 70\%. Furthermore, NL-to-DSL is robust with F1 scores greater than 35\% for queries comprising different code constructs and with different query complexities.
\end{tcolorbox}

\subsection{\ref{rq-comparison} Performance of NL-to-DSL approach against baselines}

We compare the NL-to-DSL translation approach against the baselines based on LLM and vector search. These results are tabulated in Table~\ref{tbl:baseline-table}. \\ 

\noindent \textit{Comparison against LLM Baseline.}
Table \ref{tbl:baseline-table} shows NL-to-DSL outperforms a pure LLM based baseline by 14\% on F1 score over GQL-Lite benchmark and 6\% on F1 score over Semgrep-Lite benchmark. 

\begin{figure}[h]
    \centering
    \includegraphics[width=0.9\textwidth]{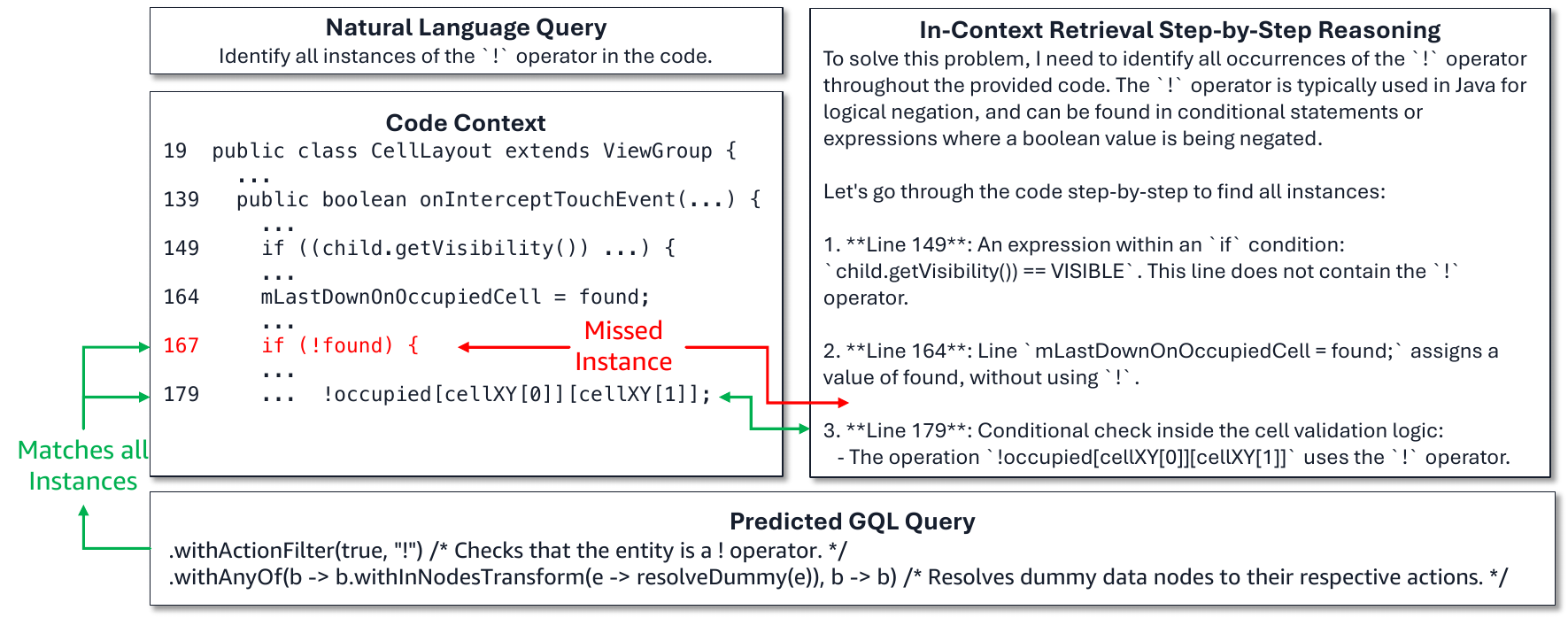}
    \caption{Code search query that is accurately answered by NL-to-GQL, but for which LLM baseline fails to identify all matches. In its step-by-step reasoning, the LLM fails to consider all instances of the negation operator in the source code.}
    \label{fig:in-context-failure}
\end{figure}

While the LLM baseline can be nearly as precise as the NL-to-DSL translation approach, they achieve significantly lower recall.  
A common failure of the LLM baseline occurs when the structural search query has multiple matches within the same method or file. In such a scenario, the LLM tends to miss some occurrences of the matching code. Figure \ref{fig:in-context-failure} shows an example of such a structural search query. Here, our approach synthesizes a GQL query that correctly matches all instances of the negation operator, whereas the LLM baseline misses an instance on line 167 in its step-by-step reasoning.

Another significant data point is that while the NL-to-DSL approach is more performant at structural code search, it is also more token efficient. Quantitatively, NL-to-DSL requires ~$22$x fewer input and output tokens as compared to the pure LLM baseline. The cost of searching code by directly calling an LLM with the input code is therefore more costly by an order of magnitude. This makes the LLM approach nonviable for search applications involving large number of repositories. This is the main reason why we report comparisons against the LLM baseline on the much smaller GQL-Lite and Semgrep-Lite benchmarks (and not on the GQL-Full and Semgrep-Full benchmarks). \\

% Furthermore, our approach can achieve $1.7\times$ higher recall with the smaller GPT-4o-mini model than In-Context Retrieval with the larger GPT-4o model.
% We see the same trend with the Semgrep-derived lite benchmark in which our structural code search approach achieves $1.1\times$ higher F1 score than In-Context Retrieval.

% \usepackage{booktabs}Structural Search

\begin{table}
\centering
\caption{Performance of NL-to-DSL approach against baselines.}
\label{tbl:baseline-table}
\begin{tabular}{llllrrr} 
\toprule \multirow{2}{*}{Benchmark} & Granularity & \multirow{2}{*}{Method}              & \multirow{2}{*}{Model}           & \multirow{2}{*}{Rec. (\%)} & \multirow{2}{*}{Prec. (\%)} & \multirow{2}{*}{F1 (\%)} \\ 
& of code match  &&&&& \\

\midrule
GQL-Lite     & Line        & NL-to-DSL & GPT-4o~~~~~~~~~ & 72.7      & 61.6       & 66.7     \\
%GQL-Lite     & Line        & Structural Search~~~~ & GPT-4o-mini~~~~ & 72.7      & 51.2       & 60.1     \\
GQL-Lite     & Line        & LLM~ & GPT-4o~~~~~~~~~ & 41.3      & 70.8       & 52.2     \\
%GQL-Lite     & Line        & In-Context Retrieval~ & GPT-4o-mini~~~~ & 14.4      & 34.5       & 20.3     \\
\midrule
Semgrep-Lite & Line        & NL-to-DSL~~~~ & GPT-4o~~~~~~~~~ & 64.0      & 52.8       & 57.9     \\
%Semgrep-Lite & Line        & Structural Search~~~~ & GPT-4o-mini~~~~ & 89.8      & 68.5       & 77.7     \\
Semgrep-Lite & Line        & LLM~ & GPT-4o~~~~~~~~~ & 56.5      & 47.0       & 51.3     \\
%Semgrep-Lite & Line        & In-Context Retrieval~ & GPT-4o-mini~~~~ & 49.1      & 16.1       & 24.3     \\
\midrule \midrule
GQL-Full     & Method      & NL-to-DSL~~~~ & GPT-4o~~~~~~~~~ & 67.7      & 65.4       & 66.5     \\
%GQL-Full     & Method      & Structural Search~~~~ & GPT-4o-mini~~~~ & 56.8      & 53.8       & 55.3     \\
GQL-Full     & Method      & V. Search T=0.25  & NV-Embed-V2     & 50.6      & 6.2        & 11.0     \\
GQL-Full     & Method      & V. Search T=0.5   & NV-Embed-V2     & 10.8      & 4.9        & 6.7      \\
GQL-Full     & Method      & V. Search T=0.75  & NV-Embed-V2     & 0.0       & 0.0        & 0.0      \\
\midrule
Semgrep-Full & Method      & NL-to-DSL~~~~ & GPT-4o~~~~~~~~~ & 71.6      & 70.5       & 71.1     \\
%Semgrep-Full & Method      & Structural Search~~~~ & GPT-4o-mini~~~~ & 66.1      & 66.4       & 66.2     \\
Semgrep-Full & Method      & V. Search T=0.25  & NV-Embed-V2     & 71.0      & 1.3        & 2.6      \\
Semgrep-Full & Method      & V. Search T=0.5   & NV-Embed-V2     & 20.1      & 10.5       & 13.8     \\
Semgrep-Full & Method      & V. Search T=0.75  & NV-Embed-V2     & 0.0       & 0.0        & 0.0      \\
\bottomrule
\end{tabular}
\end{table}

\noindent \textit{Comparison against Vector Search Baseline.}
In Table \ref{tbl:baseline-table}, we compare the performance against a vector search baseline with three different cosine similarity thresholds: T = 0.25, 0.5 and 0.75. For this comparison, we match code at a method granularity since the vector search approach retrieves method chunks. The vector search approach has a very low precision with values ranging below 11\% across all thresholds. This suggests that vector search is not capable of the complex reasoning required to precisely interpret the structural code search queries. In comparison, NL-to-DSL outperforms vector search by 55\% - 57\% on F1 scores over the GQL-Full and Semgrep-Full benchmarks.

%\begin{table}[h]
%\centering
%\begin{tabular}{lrrr} 
%\toprule
%\textbf{Baseline}                        & \textbf{Recall (\%)} & \textbf{Precision} (\%) & \textbf{F1 Score (\%) }  \\ 
%\midrule
%NL to GQL (GPT-4o-2024-08-06)            & 67.7& 65.4& 66.5\\
%NL to GQL (GPT-4o-mini)                  & 56.8& 53.8& 55.3\\
%V. Search (NV-Embed-v2) T=0.25 & 50.6& 6.2& 11.0\\
%V. Search (NV-Embed-v2) T=0.50 & 10.8& 4.9& 6.7\\
%V. Search (NV-Embed-v2) T=0.75 & 0.0& 0.0& 0.0\\
%\bottomrule
%\end{tabular}
%\caption{Method-level accuracy of code search tools on the \textbf{full GQL-derived benchmark.}}
%\label{tbl:method-level-gql}
%\end{table}
%
%\begin{table}[h]
%\centering
%\begin{tabular}{lrrr} 
%\toprule
%\textbf{Baseline}                        & \textbf{Recall (\%)} & \textbf{Precision} (\%) & \textbf{F1 Score (\%) }  \\ 
%\midrule
%NL to Semgrep (GPT-4o-2024-08-06)            & 71.6& 70.5& 71.1\\
%NL to Semgrep (GPT-4o-mini)                  & 66.1& 66.4& 66.2\\
%V. Search (NV-Embed-v2) T=0.25 & 71.0& 1.3& 2.6\\
%V. Search (NV-Embed-v2) T=0.50 & 20.1& 10.5& 13.8\\
%V. Search (NV-Embed-v2) T=0.75 & 0.0& 0.0& 0.0\\
%\bottomrule
%\end{tabular}
%\caption{Method-level accuracy of code search tools on the \textbf{full Semgrep-derived benchmark}.}
%\label{tbl:method-level-semgrep}
%\end{table}

\begin{tcolorbox}
The vector search baseline reports a very low precision for different similarity thresholds. The LLM baseline reports a significantly lower recall. In comparison, NL-to-DSL approach combines high precision with higher recall and outperforms vector search by up to 57\% and the LLM baseline by up to 14\% on F1 score. Significantly, NL-to-DSL is more token efficient by an order of magnitude against the LLM baseline.
\end{tcolorbox}

%% file: sections/RQs/rq-ablation.tex
\subsection{\ref{rq-ablation} Ablation study to determine performance contribution of RAG and query refinement}

\begin{table}[h]
\centering
\begin{tabular}{llllrrr} 
\toprule
 w/ (NL, DSL)  &w/ API  &w/ Inline  &               w/ Query  & \multirow{2}{*}{Rec. (\%)} & \multirow{2}{*}{Prec. (\%)} & \multirow{2}{*}{F1 (\%)}  \\
 Examples & docs & DSL comments & refinement \\
 
\midrule
No&Yes&No&No& 2.5          & 2.3             & 2.4             \\
  Yes&No&No&No& 34.7         & 32.4            & 33.5            \\
  Yes&Yes&No&No& 36.4         & 34.1            & 35.2            \\
  Yes&No&Yes&No& 46.3         & 44.9            & 45.6            \\
  Yes&No&Yes&Yes& 48.9         & 45.4            & 47.1            \\
  Yes&Yes&Yes&Yes& 44.5         & 40.9            & 42.6            \\
\bottomrule
\end{tabular}
\caption{Ablation study on the GQL-Full benchmark with GPT-4o-mini as the base LLM. Recall, precision and F1 scores are reported for line level code matching. }
\label{tbl:ablation}
\end{table}

In this section, we report results from the ablation study we conducted to evaluate the contribution of different components in our proposed NL-to-DSL approach. We tabulate the performance breakdown in Table~\ref{tbl:ablation}. 
Note, all experiments in this table use the same system prompt for the NL-to-GQL translation. This prompt contains an explanation of GQL and a template that outlines the requirements for a valid GQL query. We conduct ablations to evaluate four different system configurations: 
\begin{itemize}
\item[1.] RAG with few-shot examples of paired (NL, DSL) queries (column 1 in Table~\ref{tbl:ablation})
\item[2.] RAG with documentation of the DSL constructs / APIs (column 2 in Table~\ref{tbl:ablation})
\item[3.] Augmenting DSL queries in the paired (NL, DSL) examples with inline comments explaining the construction of the DSL query (column 3 in Table~\ref{tbl:ablation})
\item[4.] Query refinement using automated error detection and re-prompting (column 4 in Table~\ref{tbl:ablation})
\end{itemize}

With RAG index that is instantiated with documentation of the DSL API / constructs, LLM-to-GQL achieves a very low F1 score: 2.4\% (first row in Table~\ref{tbl:ablation}). 
In comparison, instantiating the RAG with paired (NL, DSL) queries increases the performance of the NL-to-GQL approach to 33\% F1 score. This shows that few shot examples of synthetically generated paired (NL, DSL) queries are critical for effective translation of the NL queries to DSL. Augmenting the (NL, DSL) queries with API documentation marginally increases the F1 score to 35\%.
% \ben{, but impairs performance when combined with re-prompting. Since documentation can add up to $X\times$ more tokens to the context, it is not unexpected that performance decreases as LLM reasoning capabilities decrease in long-context settings \cite{TODO}.}
On the other hand, we observe that annotating the DSL queries in the paired (NL, DSL) examples with inline comments explaining the construction of the DSL query significantly increases the performance from 33\% (in second row) to 45\% (in fourth row). An example of such inline comments can be seen in the GQL queries shown in Figure~\ref{fig:gql-query-examples}. 
This is not surprising. 
%Unlike the API documentation, these annotations are in-line comments added to the GQL component of the few-shot examples. . %In contrast, providing both API documentation and the few-shot examples only improves performance by 2\%. 
Inline comments provide explanation for each predicate in the DSL query locally, as opposed to having the LLM match the predicates in the DSL query to its corresponding documentation.
%(that explains the usage of the API along with the effects of passing arguments to the API)simultaneously reason about the effects of the arguments passed to the API (e.g. passing transitive=true, etc.).
Finally, we achieve the best system configuration with query refinement using automated error detection and re-prompting. Including this component further improves the F1 score by 1.5\% to the overall score 47\%.

\begin{tcolorbox}
Our choice of the RAG index with paired (NL, DSL) queries where the DSL queries are well annotated with comments explaining their construction, and the query refinement module are both well motivated by an increase in the overall performance on code search. 
\end{tcolorbox}

%% file: sections/RQs/rq-breakdown.tex
\subsection{Limitations of the proposed NL-to-DSL approach}

% old table
%\begin{table}[h]
%\centering
%\begin{tabular}{lcc} 
%\toprule
%& \multicolumn{2}{c}{Benchmark (Recall/Precision/F1 Score)} \\
%\cmidrule{2-3}
% & GQL-Full& Semgrep-Full  \\
% % & GQL-Full& Semgrep-Full & Average \\
%\midrule
%% NL-to-GQL& 59.9/57.3/58.5& 16.7/14.6/15.6& 38.3/35.9/37.1\\
%% NL-to-Semgrep& 26.4/30.7/28.4& 70.6/69.4/70.0& 48.5/50.1/49.2\\
%NL-to-GQL& 59.9/57.3/58.5& 16.7/14.6/15.6\\
%NL-to-Semgrep& 26.4/30.7/28.4& 70.6/69.4/70.0 \\
%\bottomrule
%\end{tabular}
%\caption{Performance of NL-to-DSL instantiated with GQL and Semgrep on the GQL-Full and Semgrep-Full benchmarks. GPT-4o is used as the base LLM and code is matched at line-level.}
%\label{tbl:cross-performance}
%\end{table}

\begin{table}[h]
\centering
\begin{tabular}{lcccccc} 
\toprule
& \multicolumn{3}{c}{GQL-Full}& \multicolumn{3}{c}{Semgrep-Full  }\\
\cmidrule(lr){2-4} \cmidrule(lr){5-7}
 & Rec. (\%)& Prec. (\%)&F1 (\%)& Rec. (\%)&Prec. (\%)& F1 (\%)\\
 % & GQL-Full& Semgrep-Full & Average \\
\midrule
% NL-to-GQL& 59.9/57.3/58.5& 16.7/14.6/15.6& 38.3/35.9/37.1\\
% NL-to-Semgrep& 26.4/30.7/28.4& 70.6/69.4/70.0& 48.5/50.1/49.2\\
NL-to-GQL& 59.9& 57.3&58.5& 16.7&14.6& 15.6\\
NL-to-Semgrep& 26.4& 30.7&28.4& 70.6&69.4& 70.0\\
\bottomrule
\end{tabular}
\caption{Performance of NL-to-DSL instantiated with GQL and Semgrep on the GQL-Full and Semgrep-Full benchmarks. GPT-4o is used as the base LLM and code is matched at line-level.}
\label{tbl:cross-performance}
\end{table}

%\ben{TODO: add diagram here}

Table \ref{tbl:cross-performance} shows the cross-benchmark scores of the best-performing configurations of NL-to-DSL instantiated with GQL and Semgrep DSLs. Each solution performs best on the benchmark derived from the same DSL. However, when tested on the opposing benchmark, NL-to-GQL's performance drops by 43\% on F1 score and and NL-to-Semgrep's performance drops by 41\% on F1 score. 
A key factor contributing to the drop in performance is the fact that GQL and Semgrep do not support the same set of code constructs and predicates.
For instance, Semgrep queries may match entire classes or method declarations whereas GQL queries can only match code constructs within a method body. 
Conversely, GQL supports interleaved data-flow and control-flow predicates that are not supported by Semgrep. 
This suggests that one may achieve the best of all worlds by building a system that combines multiple NL-to-DSL engines and a router that can route the user's natural language search queries to the most appropriate engine. 

On a different note, in this work we focus on structural code search queries that can be expressed as  conjunctions of different code constructs. Our dataset does not include queries expressed using disjunctions or negations of code constructs. However, this is not a fundamental limitation and the approach described in this paper can be extended to support such queries. 

%% file: sections/15-threats-to-validity.tex
\subsection{Threats to the validity of results}

The algorithm that generates the paired (NL, DSL) queries is used both to prepare the evaluation dataset as well as to generate the few-shot examples used in the RAG setup. To ensure there is no leakage from the RAG examples to the evaluation dataset, we check that the code corpus used for benchmarks and the RAG examples are completely disjoint. 
Further, we filter out duplicate queries between the RAG examples and the evaluation dataset by matching the DSL queries.
We identified such duplicates for simple single-term queries such as \textit{"Find all continue statements"}. 

% We tried to be exhaustive in terms of supported construct types, but we only support a small subset of all the available GQL predicates.
% Similarly for Semgrep, we did not explore the taint-analysis mode or typed metavariables.

%% file: sections/16-related-work.tex
\section{Related Work}
\label{sec:related-work}

\noindent \textbf{Structural Code Search:}
Structural code search tools enable users to search for code based on its syntactic structure.
Comby\cite{comby}, Semgrep~\cite{semgrep}, CodeQL~\cite{codeql} and GQL~\cite{mukherjee-gql} are examples of different DSLs that enable structural code search. 
Tools that accompany these DSLs-- Semgrep, CodeQL, and GQL are most commonly used to detect potential vulnerabilities, code anti-patterns, and bugs by executing a pre-written database of structural search queries over a developer's code-base.
Structural code search capabilities are also present in modern IDEs such as IntelliJ~\cite{intellij} or IDE plugins such as CodeQue Visual Studio Extension~\cite{codeque}.
%IntelliJ provides a user interface to assist developers in manually constructing structural code search queries~\cite{intellij}. 
However, no current structural code search tool allows developers to express queries in natural language. \\

%IDEs like Visual Studio Code and IntelliJ also provide code navigation tools that allow developers to ...\ben{are code navigation tools related work? e.g. go to definition, go to reference, etc.}

\noindent \textbf{Semantic Code Search:}
Semantic code search is the task of retrieving relevant code to answer natural language queries using code embeddings.
CodeSearchNet is a semantic code search benchmark which mines natural language queries and their matching code by extracting function docstrings and comments~\cite{codesearchnet}.
The Neural-Code-Search-Evaluation-Dataset (NCSE) is another semantic code search benchmark which mines StackOverflow questions and answers for natural language queries and matching code~\cite{neuralcodesearch}.
These benchmarks differ from our structural code search benchmark in that their natural language queries describe the high-level functionality or the semantics of the code rather than its structure or specific implementation. As an example, the query \textit{"Sending an Intent to browser to open specific URL"} from the NCSE dataset describes the intent / functionality of the code.

\noindent \textbf{In-Context Retrieval:}
Needle-in-a-haystack style benchmarks measure the ability of an LLM to retrieve specific information from its context~\cite{needle-in-haystack}.
Such benchmarks embed a random fact or statement within a large string of text designed to distract from the embedded information. The LLM is then tested on its ability to recall the embedded fact from its context.
The needle-in-a-haystack task is similar to the task we pose to the LLM retrieval baseline in that they both ask an LLM to answer a query by retrieving information (or code in our case) from the LLM's context.
RepoQA is a related benchmark that tests an LLM's ability to accurately retrieve functions from long context given a description of the function~\cite{repoqa}.

Our benchmark differs from these needle-in-a-haystack benchmarks in that our structural code search queries require the LLM to reason about the structure of the code to be retrieved.
Additionally, our benchmark requires baselines to search through much larger code corpus (>4MB in size) than would fit in any LLM context window. 
We also maximize the performance of the In-Context Retrieval baselines by limiting the length of code provided to the LLM by prompting the LLM to answer each query with respect to each Java class separately. \\

%Additionally, in our benchmark, the target of the queries sourced from real-world Java projects.

\noindent \textbf{LLM Coding Assistants:}
LLM-based coding assistants such as Github's Copilot~\cite{copilot}, Codeium~\cite{codeium}, Amazon Q Developer~\cite{amazonqdev}, Tabnine~\cite{tabnine}, and more enable developers to pose natural language queries in a chat interface. These tools index the code-base to allow an LLM to retrieve specific source files on demand to answer the user's query.
While we do not directly test these coding assistants, we approximate their functionality with the LLM baseline in our evaluation.

%% file: sections/17-conclusion.tex
\section{Conclusions}
\label{sec:conc}

Code search is an integral part of software developers' daily workflow.
Structural code search promises to enrich search capabilities by enabling more expressive queries for a variety of developer tasks such as refactoring, code navigation, and bug-finding.
Yet adoption for structural code search engines is low as they require users to learn domain-specific languages to express their queries.
In this work, we introduced a novel approach to enable developers to express structural code search queries in natural language.
By lowering the barrier to entry, our approach empowers developers with structural code search, and promises to enable new use cases and greater productivity.

% \noindent \textbf{Future Work}
% Currently, our implementation only considers conjunctions of GQL predicates and Semgrep clauses. We plan to extend our approach to include support for queries with disjunctions and negations. As our implementation is easily extensible, we also plan to add support for additional GQL predicates.

% To close this gap, our work proposes a new approach to provide users the ease of use and flexibility of expressing structural code search queries in natural language.
% By combining the reasoning capabilities of an LLM with RAG with a structural code search engine, our approach more accurately and efficiently answers structural search queries.

%% file: main.bbl

\begin{thebibliography}{24}

%%% the Web address.

\ifx \showCODEN    \undefined \def \showCODEN     #1{\unskip}     \fi
\ifx \showDOI      \undefined \def \showDOI       #1{#1}\fi
\ifx \showISBNx    \undefined \def \showISBNx     #1{\unskip}     \fi
\ifx \showISBNxiii \undefined \def \showISBNxiii  #1{\unskip}     \fi
\ifx \showISSN     \undefined \def \showISSN      #1{\unskip}     \fi
\ifx \showLCCN     \undefined \def \showLCCN      #1{\unskip}     \fi
\ifx \shownote     \undefined \def \shownote      #1{#1}          \fi
\ifx \showarticletitle \undefined \def \showarticletitle #1{#1}   \fi
\ifx \showURL      \undefined \def \showURL       {\relax}        \fi
% invisible to TeX
\providecommand\bibfield[2]{#2}
\providecommand\bibinfo[2]{#2}
\providecommand\natexlab[1]{#1}
\providecommand\showeprint[2][]{arXiv:#2}

\bibitem[Amazon Web~Services(2024)]%
        {amazonqdev}
\bibfield{author}{\bibinfo{person}{Inc. Amazon Web~Services}.} \bibinfo{year}{2024}\natexlab{}.
\newblock \bibinfo{title}{AI For Software Developement - Amazon Q Developer - AWS}.
\newblock
\newblock
\urldef\tempurl%
\url{https://aws.amazon.com/q/developer/}
\showURL{%
\tempurl}


\bibitem[Bajracharya and Lopes(2010)]%
        {Bajracharya-queries-are-short}
\bibfield{author}{\bibinfo{person}{Sushil~Krishna Bajracharya} {and} \bibinfo{person}{Cristina~Videira Lopes}.} \bibinfo{year}{2010}\natexlab{}.
\newblock \showarticletitle{Analyzing and mining a code search engine usage log}.
\newblock \bibinfo{journal}{\emph{Empirical Software Engineering}} \bibinfo{volume}{17}, \bibinfo{number}{4–5} (\bibinfo{date}{Sept.} \bibinfo{year}{2010}), \bibinfo{pages}{424–466}.
\newblock
\showISSN{1573-7616}
\urldef\tempurl%
\url{https://doi.org/10.1007/s10664-010-9144-6}
\showDOI{\tempurl}


\bibitem[Cambronero et~al\mbox{.}(2019)]%
        {cambronero-code-search}
\bibfield{author}{\bibinfo{person}{Jose Cambronero}, \bibinfo{person}{Hongyu Li}, \bibinfo{person}{Seohyun Kim}, \bibinfo{person}{Koushik Sen}, {and} \bibinfo{person}{Satish Chandra}.} \bibinfo{year}{2019}\natexlab{}.
\newblock \showarticletitle{When deep learning met code search}. In \bibinfo{booktitle}{\emph{Proceedings of the 2019 27th ACM Joint Meeting on European Software Engineering Conference and Symposium on the Foundations of Software Engineering}} (Tallinn, Estonia) \emph{(\bibinfo{series}{ESEC/FSE 2019})}. \bibinfo{publisher}{Association for Computing Machinery}, \bibinfo{address}{New York, NY, USA}, \bibinfo{pages}{964–974}.
\newblock
\showISBNx{9781450355728}
\urldef\tempurl%
\url{https://doi.org/10.1145/3338906.3340458}
\showDOI{\tempurl}


\bibitem[Codeium(2024)]%
        {codeium}
\bibfield{author}{\bibinfo{person}{Inc. Codeium}.} \bibinfo{year}{2024}\natexlab{}.
\newblock \bibinfo{title}{Codeium - AI Code Completion and Chat}.
\newblock
\newblock
\urldef\tempurl%
\url{https://codeium.com/}
\showURL{%
\tempurl}


\bibitem[CodeQue.co(2024)]%
        {codeque}
\bibfield{author}{\bibinfo{person}{CodeQue.co}.} \bibinfo{year}{2024}\natexlab{}.
\newblock \bibinfo{title}{Multiline \& Structural Code Search}.
\newblock
\newblock
\urldef\tempurl%
\url{https://marketplace.visualstudio.com/items?itemName=CodeQue.codeque}
\showURL{%
\tempurl}


\bibitem[Di~Grazia and Pradel(2023)]%
        {pradel-code-search-survey}
\bibfield{author}{\bibinfo{person}{Luca Di~Grazia} {and} \bibinfo{person}{Michael Pradel}.} \bibinfo{year}{2023}\natexlab{}.
\newblock \showarticletitle{Code Search: A Survey of Techniques for Finding Code}.
\newblock \bibinfo{journal}{\emph{ACM Comput. Surv.}} \bibinfo{volume}{55}, \bibinfo{number}{11}, Article \bibinfo{articleno}{220} (\bibinfo{date}{Feb.} \bibinfo{year}{2023}), \bibinfo{numpages}{31}~pages.
\newblock
\showISSN{0360-0300}
\urldef\tempurl%
\url{https://doi.org/10.1145/3565971}
\showDOI{\tempurl}


\bibitem[Github(2024a)]%
        {codeql}
\bibfield{author}{\bibinfo{person}{Inc. Github}.} \bibinfo{year}{2024}\natexlab{a}.
\newblock \bibinfo{title}{CodeQL overview - CodeQL}.
\newblock
\newblock
\urldef\tempurl%
\url{https://codeql.github.com/docs/codeql-overview/}
\showURL{%
\tempurl}


\bibitem[Github(2024b)]%
        {copilot}
\bibfield{author}{\bibinfo{person}{Inc. Github}.} \bibinfo{year}{2024}\natexlab{b}.
\newblock \bibinfo{title}{Github Copilot}.
\newblock
\newblock
\urldef\tempurl%
\url{https://github.com/features/copilot}
\showURL{%
\tempurl}


\bibitem[Husain et~al\mbox{.}(2020)]%
        {codesearchnet}
\bibfield{author}{\bibinfo{person}{Hamel Husain}, \bibinfo{person}{Ho-Hsiang Wu}, \bibinfo{person}{Tiferet Gazit}, \bibinfo{person}{Miltiadis Allamanis}, {and} \bibinfo{person}{Marc Brockschmidt}.} \bibinfo{year}{2020}\natexlab{}.
\newblock \bibinfo{title}{CodeSearchNet Challenge: Evaluating the State of Semantic Code Search}.
\newblock
\newblock
\showeprint[arxiv]{1909.09436}~[cs.LG]
\urldef\tempurl%
\url{https://arxiv.org/abs/1909.09436}
\showURL{%
\tempurl}


\bibitem[JetBrains(2024)]%
        {intellij}
\bibfield{author}{\bibinfo{person}{JetBrains}.} \bibinfo{year}{2024}\natexlab{}.
\newblock \bibinfo{title}{Structural search and replace}.
\newblock
\newblock
\urldef\tempurl%
\url{https://www.jetbrains.com/help/idea/structural-search-and-replace.html}
\showURL{%
\tempurl}


\bibitem[Kamradt(2023)]%
        {needle-in-haystack}
\bibfield{author}{\bibinfo{person}{Gregory Kamradt}.} \bibinfo{year}{2023}\natexlab{}.
\newblock \bibinfo{title}{Needle In A Haystack - Pressure Testing LLMs}.
\newblock
\newblock
\urldef\tempurl%
\url{https://github.com/gkamradt/LLMTest_NeedleInAHaystack/blob/main/README.md}
\showURL{%
\tempurl}


\bibitem[Lawall(2023)]%
        {coccinelle-origins}
\bibfield{author}{\bibinfo{person}{Julia Lawall}.} \bibinfo{year}{2023}\natexlab{}.
\newblock \showarticletitle{{On the Origins of Coccinelle}}. In \bibinfo{booktitle}{\emph{Eelco Visser Commemorative Symposium (EVCS 2023)}} \emph{(\bibinfo{series}{Open Access Series in Informatics (OASIcs)}, Vol.~\bibinfo{volume}{109})}, \bibfield{editor}{\bibinfo{person}{Ralf L\"{a}mmel}, \bibinfo{person}{Peter~D. Mosses}, {and} \bibinfo{person}{Friedrich Steimann}} (Eds.). \bibinfo{publisher}{Schloss Dagstuhl -- Leibniz-Zentrum f{\"u}r Informatik}, \bibinfo{address}{Dagstuhl, Germany}, \bibinfo{pages}{18:1--18:11}.
\newblock
\showISBNx{978-3-95977-267-9}
\showISSN{2190-6807}
\urldef\tempurl%
\url{https://doi.org/10.4230/OASIcs.EVCS.2023.18}
\showDOI{\tempurl}


\bibitem[Lewis et~al\mbox{.}(2020)]%
        {lewis-rag}
\bibfield{author}{\bibinfo{person}{Patrick Lewis}, \bibinfo{person}{Ethan Perez}, \bibinfo{person}{Aleksandra Piktus}, \bibinfo{person}{Fabio Petroni}, \bibinfo{person}{Vladimir Karpukhin}, \bibinfo{person}{Naman Goyal}, \bibinfo{person}{Heinrich K\"{u}ttler}, \bibinfo{person}{Mike Lewis}, \bibinfo{person}{Wen-tau Yih}, \bibinfo{person}{Tim Rockt\"{a}schel}, \bibinfo{person}{Sebastian Riedel}, {and} \bibinfo{person}{Douwe Kiela}.} \bibinfo{year}{2020}\natexlab{}.
\newblock \showarticletitle{Retrieval-augmented generation for knowledge-intensive NLP tasks}. In \bibinfo{booktitle}{\emph{Proceedings of the 34th International Conference on Neural Information Processing Systems}} (Vancouver, BC, Canada) \emph{(\bibinfo{series}{NIPS '20})}. \bibinfo{publisher}{Curran Associates Inc.}, \bibinfo{address}{Red Hook, NY, USA}, Article \bibinfo{articleno}{793}, \bibinfo{numpages}{16}~pages.
\newblock
\showISBNx{9781713829546}


\bibitem[Li et~al\mbox{.}(2019)]%
        {neuralcodesearch}
\bibfield{author}{\bibinfo{person}{Hongyu Li}, \bibinfo{person}{Seohyun Kim}, {and} \bibinfo{person}{Satish Chandra}.} \bibinfo{year}{2019}\natexlab{}.
\newblock \bibinfo{title}{Neural Code Search Evaluation Dataset}.
\newblock
\newblock
\showeprint[arxiv]{1908.09804}~[cs.SE]
\urldef\tempurl%
\url{https://arxiv.org/abs/1908.09804}
\showURL{%
\tempurl}


\bibitem[Liu et~al\mbox{.}(2024)]%
        {repoqa}
\bibfield{author}{\bibinfo{person}{Jiawei Liu}, \bibinfo{person}{Jia~Le Tian}, \bibinfo{person}{Vijay Daita}, \bibinfo{person}{Yuxiang Wei}, \bibinfo{person}{Yifeng Ding}, \bibinfo{person}{Yuhan~Katherine Wang}, \bibinfo{person}{Jun Yang}, {and} \bibinfo{person}{Lingming Zhang}.} \bibinfo{year}{2024}\natexlab{}.
\newblock \bibinfo{title}{RepoQA: Evaluating Long Context Code Understanding}.
\newblock
\newblock
\showeprint[arxiv]{2406.06025}~[cs.SE]
\urldef\tempurl%
\url{https://arxiv.org/abs/2406.06025}
\showURL{%
\tempurl}


\bibitem[Microsoft(2024)]%
        {vscode}
\bibfield{author}{\bibinfo{person}{Microsoft}.} \bibinfo{year}{2024}\natexlab{}.
\newblock \bibinfo{title}{Visual Studio Code - Code Navigation}.
\newblock
\newblock
\urldef\tempurl%
\url{https://code.visualstudio.com/docs/editor/editingevolved}
\showURL{%
\tempurl}


\bibitem[Mukherjee et~al\mbox{.}(2022)]%
        {mukherjee-gql}
\bibfield{author}{\bibinfo{person}{Rajdeep Mukherjee}, \bibinfo{person}{Omer Tripp}, \bibinfo{person}{Ben Liblit}, {and} \bibinfo{person}{Michael Wilson}.} \bibinfo{year}{2022}\natexlab{}.
\newblock \showarticletitle{{Static Analysis for AWS Best Practices in Python Code}}. In \bibinfo{booktitle}{\emph{36th European Conference on Object-Oriented Programming (ECOOP 2022)}} \emph{(\bibinfo{series}{Leibniz International Proceedings in Informatics (LIPIcs)}, Vol.~\bibinfo{volume}{222})}, \bibfield{editor}{\bibinfo{person}{Karim Ali} {and} \bibinfo{person}{Jan Vitek}} (Eds.). \bibinfo{publisher}{Schloss Dagstuhl -- Leibniz-Zentrum f{\"u}r Informatik}, \bibinfo{address}{Dagstuhl, Germany}, \bibinfo{pages}{14:1--14:28}.
\newblock
\showISBNx{978-3-95977-225-9}
\showISSN{1868-8969}
\urldef\tempurl%
\url{https://doi.org/10.4230/LIPIcs.ECOOP.2022.14}
\showDOI{\tempurl}


\bibitem[Sadowski et~al\mbox{.}(2015)]%
        {sadowski-code-search}
\bibfield{author}{\bibinfo{person}{Caitlin Sadowski}, \bibinfo{person}{Kathryn~T. Stolee}, {and} \bibinfo{person}{Sebastian Elbaum}.} \bibinfo{year}{2015}\natexlab{}.
\newblock \showarticletitle{How developers search for code: a case study}. In \bibinfo{booktitle}{\emph{Proceedings of the 2015 10th Joint Meeting on Foundations of Software Engineering}} (Bergamo, Italy) \emph{(\bibinfo{series}{ESEC/FSE 2015})}. \bibinfo{publisher}{Association for Computing Machinery}, \bibinfo{address}{New York, NY, USA}, \bibinfo{pages}{191–201}.
\newblock
\showISBNx{9781450336758}
\urldef\tempurl%
\url{https://doi.org/10.1145/2786805.2786855}
\showDOI{\tempurl}


\bibitem[Semgrep(2024)]%
        {semgrep}
\bibfield{author}{\bibinfo{person}{Inc. Semgrep}.} \bibinfo{year}{2024}\natexlab{}.
\newblock \bibinfo{title}{Semgrep | Homepage}.
\newblock
\newblock
\urldef\tempurl%
\url{https://semgrep.dev/}
\showURL{%
\tempurl}


\bibitem[Sim et~al\mbox{.}(2011)]%
        {Sim-queries-are-short}
\bibfield{author}{\bibinfo{person}{Susan~Elliott Sim}, \bibinfo{person}{Medha Umarji}, \bibinfo{person}{Sukanya Ratanotayanon}, {and} \bibinfo{person}{Cristina~V. Lopes}.} \bibinfo{year}{2011}\natexlab{}.
\newblock \showarticletitle{How Well Do Search Engines Support Code Retrieval on the Web?}
\newblock \bibinfo{journal}{\emph{ACM Transactions on Software Engineering and Methodology}} \bibinfo{volume}{21}, \bibinfo{number}{1} (\bibinfo{date}{Dec.} \bibinfo{year}{2011}), \bibinfo{pages}{1–25}.
\newblock
\showISSN{1557-7392}
\urldef\tempurl%
\url{https://doi.org/10.1145/2063239.2063243}
\showDOI{\tempurl}


\bibitem[SourceGraph(2024)]%
        {sourcegraph}
\bibfield{author}{\bibinfo{person}{SourceGraph}.} \bibinfo{year}{2024}\natexlab{}.
\newblock \bibinfo{title}{SourceGraph - Code Search}.
\newblock
\newblock
\urldef\tempurl%
\url{https://sourcegraph.com/code-search}
\showURL{%
\tempurl}


\bibitem[Svajlenko and Roy(2015)]%
        {bigclonebench-ijadataset}
\bibfield{author}{\bibinfo{person}{Jeffrey Svajlenko} {and} \bibinfo{person}{Chanchal~K. Roy}.} \bibinfo{year}{2015}\natexlab{}.
\newblock \showarticletitle{Evaluating clone detection tools with BigCloneBench}. In \bibinfo{booktitle}{\emph{2015 IEEE International Conference on Software Maintenance and Evolution (ICSME)}}. \bibinfo{pages}{131--140}.
\newblock
\urldef\tempurl%
\url{https://doi.org/10.1109/ICSM.2015.7332459}
\showDOI{\tempurl}


\bibitem[Tabnine(2024)]%
        {tabnine}
\bibfield{author}{\bibinfo{person}{Inc. Tabnine}.} \bibinfo{year}{2024}\natexlab{}.
\newblock \bibinfo{title}{Tabnine AI Code Assistant}.
\newblock
\newblock
\urldef\tempurl%
\url{https://www.tabnine.com/}
\showURL{%
\tempurl}


\bibitem[van Tonder(2024)]%
        {comby}
\bibfield{author}{\bibinfo{person}{Rijnard van Tonder}.} \bibinfo{year}{2024}\natexlab{}.
\newblock \bibinfo{title}{Comby - Structural code search and replace for every language}.
\newblock
\newblock
\urldef\tempurl%
\url{https://comby.dev/}
\showURL{%
\tempurl}


\end{thebibliography}
